\documentclass[12pt,preprint]{emulateapj}

\begin{document}

\title{Mapping the Shores of the Brown Dwarf Desert, I: Upper Scorpius}

\author{Adam L. Kraus}
\affil{Department of Astrophysics, California Institute of Technology, Pasadena, CA 91125} 
\author{Michael J. Ireland}
\affil{Division of Geological and Planetary Sciences, California Institute of Technology, Pasadena, CA 91125}
\author{Frantz Martinache, James P. Lloyd}
\affil{Department of Astronomy, Cornell University, Ithaca, NY 14853}

\begin{abstract}

We present the results of a survey for stellar and substellar companions 
to 82 young stars in the nearby OB association Upper Scorpius. This survey 
used nonredundant aperture-mask interferometry to achieve typical contrast 
limits of $\Delta$$K\sim$5-6 at the diffraction limit, revealing 12 new 
binary companions that lay below the detection limits of traditional 
high-resolution imaging; we also summarize a complementary snapshot 
imaging survey that discovered 7 directly resolved companions. The overall 
frequency of binary companions ($\sim$35$^{+5}_{-4}$\% at separations of 
6-435 AU) appears to be equivalent to field stars of similar mass, but 
companions could be more common among lower-mass stars than for the field. 
The companion mass function has statistically significant differences 
compared to several suggested mass functions for the field, and we suggest 
an alternate log-normal parameterization of the mass-function. Our survey 
limits encompass the entire brown dwarf mass range, but we only detected a 
single companion that might be a brown dwarf; this deficit resembles the 
so-called ``brown dwarf desert'' that has been observed by radial-velocity 
planet searches. Finally, our survey's deep detection limits extend into 
the top of the planetary mass function, reaching 8-12 $M_{Jup}$ for half 
of our sample. We have not identified any planetary companions at high 
confidence ($\ga$99.5\%), but we have identified four candidate companions 
at lower confidence ($\ga$97.5\%) that merit additional followup to 
confirm or disprove their existence.

\end{abstract}

\keywords{stars:binaries:general; stars:low-mass,brown 
dwarfs;stars:pre-main sequence}

\section{Introduction}

The detection and characterization of low-mass companions has become 
one of the highest priorities of the astronomical community. Radial 
velocity surveys have discovered over 200 extrasolar planetary 
companions over the past decade, and both RV surveys and 
coronagraphic imaging surveys have discovered an abundance of 
stellar-mass companions (e.g. Marcy \& Butler 2000; McCarthy \& 
Zuckerman 2004; Metchev 2005; Johnson et al. 2006; Naef et al. 
2007). However, very few brown dwarf companions have been 
identified, an unexpected result given that the observational 
signatures of more massive companions are far larger than those of 
planetary-mass companions and that free-floating brown dwarfs are 
very common (Kirkpatrick et al. 2000; Luhman et al. 2003; Chiu et 
al. 2006; Slesnick et al. 2006a, 2006b). This dearth of companions 
between the stellar and planetary mass regimes is popularly known as 
the ``brown dwarf desert''. The existence and extent of the brown 
dwarf desert can provide key constraints on star and planet 
formation since it represents the extreme mass limit of both 
processes.

If the stellar-mass binary companions of solar-mass stars are drawn 
from the Initial Mass Function (IMF; e.g. Kroupa 1995) or formed via 
some other process that preferentially forms low-mass companions 
(e.g. Duquennoy \& Mayor 1991, hereafter DM91), then brown dwarf 
companions should be common unless another process inhibits their 
formation or dynamically strips them. However, if stellar companions 
are formed via the fragmentation of a protostellar core, then there 
are no a priori expectations that brown dwarfs should form. Indeed, 
even if fragmentation can form an extremely inequal-mass pair, the 
long collapse timescale for low-mass objects might lead to their 
preferential photoevaporation by the higher-mass, more luminous 
companion.

It is also unclear whether brown dwarfs could form via planetary 
formation processes. Radial velocity surveys suggest that the giant 
planetary mass function is well-fit by a power law, 
$dN/dm\propto$$M^{-1.05}$, for masses of $\sim$1-10 $M_{Jup}$ (Marcy 
et al. 2005). If this power law extends to higher masses, there 
should be as many ``planetary'' companions with masses of 10-25 
$M_{Jup}$ as with masses of 4-10 $M_{Jup}$ or 1.6-4.0 $M_{Jup}$. An 
absence of these companions suggests either that the function is not 
a power law or that the power law is truncated by some limit. For 
example, submillimeter disk surveys suggest that protoplanetary 
disks have a mean mass of $\sim$5 $M_{Jup}$ by the age of 1-2 Myr 
(Andrews \& Williams 2005), with a small fraction ($\sim$5\%) having 
masses of $\sim$30-100 $M_{Jup}$. Unless massive planets are formed 
very early or efficiently accrete the entire disk mass, this could 
impose an upper cutoff on the distribution of planetary masses.

The brown dwarf desert has been studied mostly at very small or very 
large separations. The radial velocity exoplanet surveys that have 
proven so successful over the past decade should have detected any 
brown dwarfs within their outer separation limit ($\sim$3-5 AU), and 
they have set very low upper limits on the frequency of close brown 
dwarf companions to solar mass stars ($<$1\%; Marcy \& Butler 2000; 
Grether \& Lineweaver 2006). Similarly, high-resolution 
coronagraphic imaging surveys have demonstrated sufficient 
sensitivity to identify brown dwarf companions at typical separation 
limits of $>$50 AU (e.g. Gizis et al. 2001; Neuh\"auser et al. 2003; 
McCarthy \& Zuckerman 2004; Neuh\"auser \& Guenther 2004; Metchev 
2005). They have measured frequencies which are low, but somewhat 
inconsistent (and perhaps not anomalously low; 1$\pm$1\% by McCarthy 
\& Zuckerman, compared to 6.8$^{+8.3}_{-4.9}$\% by Metchev and 
18$\pm$14\% by Gizis et al.). A survey for wide companions to 
high-mass (2-8 $M_{\sun}$) stars in Upper Sco by Kouwenhoven et al. 
(2007) found a relatively low frequency for brown dwarf companions, 
0.5$\pm$0.5\% at separations of 130-520 AU. Finally, there have been 
an intriguing sample of candidate planetary-mass companions 
identified at large separations (e.g. Chauvin et al. 2004; 
Neuh\"auser et al. 2005), but both their mass and formation 
mechanism are still uncertain and their frequency is still 
unconstrained (e.g. Masciadri et al. 2005; Kraus et al. 2006; Biller 
et al. 2007; Ahmic et al. 2007).

However, these surveys don't study the actual separation range where most 
giant planets and binary companions are expected to form. Most giant 
planets at small orbital radii ($\la$5 AU) are thought to have migrated 
inward, so their mass distribution may not match that of their more 
distant brethren. The binary formation process may also be different for 
small separations ($\la$10 AU), with H$_2$ dissociation softening the 
equation of state and leading to enhanced fragmentation over that expected 
for larger length scales (Whitworth \& Stamatellos 2006). Similarly, giant 
planets aren't expected to form at very large radii ($\ga$30 AU) since the 
formation timescale is too long, and the frequency of wide binary 
companions may differ significantly from those of closer binaries (e.g. 
Kraus \& Hillenbrand 2007a) since the fragmentation occurs on a length 
scale that is several orders of magnitude larger.

Ideally, the desert should be studied at the separation range where giant 
planets and most binaries are thought to form ($\sim$5-30 AU; Lissauer \& 
Stephenson 2007; DM91), but this has been impossible using existing 
techniques. For example, theoretical models (Chabrier et al. 2000) suggest 
that a 50 $M_{Jup}$ brown dwarf located 15 AU (1-2$\arcsec$) from a nearby 
field star will have a contrast ratio of $\Delta$$K\sim$10-15 magnitudes 
at a separation of only $\sim$1\arcsec. The contrast problem could have 
been addressed by observing young stars since their substellar companions 
would be intrinsically more luminous ($\Delta$$K<$5 mag), but most young 
stars are further away, so the separations are even smaller 
(0.1-0.2$\arcsec$; $\sim$$\lambda$$/D$). Sensitivities near the 
diffraction limit have traditionally been far too shallow to detect such 
companions. However, new advances in high-resolution imaging techniques 
are now opening up this critical regime; our survey will use one such 
technique, non-redundant aperture mask interferometry.

The technique of non-redundant aperture masking has been well-established
as a means of achieving the full diffraction limit of a single telescope
(e.g. Nakajima et al. 1989; Tuthill et al. 2000). The reason
for the technique's success over direct imaging is that the calibration
is independent of structure of the wavefront over scales larger than a
single sub-aperture, but it still preserves the angular resolution of
the full aperture. This technique, when applied to seeing-limited
observations, requires observations to be taken in a speckle mode with
sub-apertures of diameter smaller than the atmospheric coherence length,
limiting the technique to objects brighter than about $m_H=5$. The
use of adaptive optics allows for longer integration times and larger
sub-apertures, extending the technique to much fainter targets.

Published detections have been able to recover astrometrically discovered 
binary systems with contrast ratios of 3:1 at 0.6 $\lambda$$/D$ and 100:1 
at $\lambda$$/D$ (Pravdo et al. 2006; Lloyd et al. 2006; Ireland et
al. 2008) using total  observation times of $\sim$10 minutes. 
The inner limit of companion detectability at high contrast is
$\lambda/2B_L$, where $B_L$ is the longest baseline in the mask
(typically 80-95\% of the aperture diameter). 
Typical closure
phase errors are such that aperture masking can unveil high contrast
companions at separations ~5 times closer than direct imaging in both
$H$ and $K$ bands.  

In this paper, we describe an aperture-mask interferometry and direct 
imaging survey to detect stellar and substellar companions to young stars 
in the nearby OB association Upper Scorpius. This survey directly studies 
the age and separation range corresponding to the peak of planet 
formation, offering the first glimpse of the brown dwarf desert in this 
critical range of parameter space. In Section 2, we describe our survey 
sample, and in Section 3, we describe the observations and data analysis 
techniques. In Section 4, we summarize the results of our survey. In 
Section 5, we combine these results with previous binary surveys to place 
constraints on the stellar binary frequency, mass function, and separation 
distribution, and in Section 6, we consider constraints on the 
corresponding parameters for the planetary population. Finally, in Section 
7, we discuss the implications of our survey for the extent and aridity of 
the brown dwarf desert.

\section{Survey Sample}

Upper Sco is an ideal target for large-scale surveys to detect brown 
dwarf or planetary companions. It is young enough ($\sim$5 Myr) that 
substellar companions are much more luminous than those of typical field 
stars, and this age is thought to be the peak epoch of giant planet 
formation (Lissauer \& Stephenson 2007, and references therein). Its 
relative proximity ($\sim$145 pc; de Zeeuw et al. 1999) also means that 
the resolution limit of large telescopes ($\sim$40-100 mas; 6-15 AU) 
corresponds to the giant planet separation regime of our own solar 
system. Finally, the association has been very intensely studied, with 
several hundred members identified in the past decade, so it provides a 
much larger sample of well-characterized members than nearby moving 
groups.

We compiled a preliminary list of 356 targets from the known members 
of Upper Sco as compiled in Kraus \& Hillenbrand (2007a); this 
census included all spectroscopically-confirmed members with 
spectral type G0 or later from the recent surveys by Walter et al. 
(1994), Preibisch et al. (1998, 2001, 2002), Kunkel (1999), Ardila 
et al. (2000), Martin et al. (2004), and Slesnick et al. (2006a). We 
also added two stars that were not included in that census: 
RXJ1550.9-2534 (which was originally classified as F9 by the HD 
catalogue but was reclassified as G1 by the Michigan Spectral 
Survey; Houk \& Smith-Moore 1988) and V1149 Sco (which was not 
included in any large membership surveys since it was identified as 
a young star before they were conducted; Stephenson 1986).

All of our observations have been conducted from northern sites, so 
we removed 25 of the 26 stars south of $\delta=-25$$^o$ from further 
consideration. The only exception was RXJ1550.9-2534, which we 
retained in order to make a complete group of four science targets 
of similar brightness and airmass. As we describe below, preliminary 
imaging showed that it was an obvious binary that is not suitable 
for masking observations anyway, but we retain it in our sample for 
statistical purposes. We also rejected 230 of the remaining low-mass 
association members which were optically fainter than the useful 
limit of the AO system ($R\ga$14). Finally, we removed the 23
known binaries with separations of $<$3\arcsec. In close systems, 
the stellar companion would have dominated the signal in our 
observations, complicating any search for fainter companions. Wider 
binaries (with separations near the seeing limit) were rejected 
because they are generally not corrected well by the AO system, 
though we still observed several of them with direct imaging in 
order to test whether this would actually occur. A total of seven 
targets were not discovered to be binaries until we arrived at the 
telescope and obtained quick direct imaging observations; these 
targets were also removed from the aperture mask sample. We 
mistakenly removed one more target, USco-160643.8-190805, that we 
initially thought was a binary based on direct imaging. Subsequent 
analysis showed that it was flagged as a binary erroneously; we will 
not consider it in our statistical sample because we only have 
imaging data and not masking data.

These cuts left a total of 72 Upper Sco members in our aperture mask 
sample, plus 11 members (10 known or new binary systems and 1 erroneous 
omission) that we only observed with direct imaging. We list all of these 
targets in Table 1, where we also include each target's spectral type 
(adopted from the discovery survey), mass (as determined in Section 3.3), 
and $R$ and $K$ magnitudes, plus the target group that it was observed 
with (as described in Section 3.1). In Table 2, we list the 19 known 
binary systems that would have passed our selection criteria. We did not 
observe any of these systems, but we will include them in our sample for 
determining stellar binary statistics since they have known binary 
companions. Our upper limits on the existence of planetary-mass companions 
will not include any known or newly-discovered binary systems.

\begin{deluxetable*}{lllccrrc}
\tabletypesize{\scriptsize}
\tablewidth{0pt}
\tablecaption{Upper Sco sample}
\tablehead{\colhead{Name} & \colhead{RA} & \colhead{DEC} & 
\colhead{SpT} & \colhead{Mass} & \colhead{$R$} & \colhead{$K$} &
\colhead{Group}
\\
\colhead{} & \multicolumn{2}{c}{(J2000)} & \colhead{} & 
\colhead{($M_{\sun}$)} & \colhead{(mag)} & \colhead{(mag)}
}
\startdata
RXJ1550.0-2312&15 50 04.99&-23 11 53.7&M2&0.49&13.1&8.93&14\\
RXJ1550.9-2534&15 50 56.42&-25 34 19.0&G1&1.75&9.4&7.91&...\\
RXJ1551.1-2402&15 51 06.61&-24 02 19.0&M2&0.49&13.5&9.73&17\\
RXJ1557.8-2305&15 57 50.03&-23 05 09.4&M0&0.68&12.7&9.27&12\\
RXJ1558.1-2405&15 58 08.15&-24 05 53.0&K4&0.95&11.4&8.96&8\\
RXJ1558.2-2328&15 58 12.71&-23 28 36.4&G2&1.66&9.9&8.02&1\\
RXJ1600.2-2417&16 00 13.30&-24 18 10.6&M0&0.68&13.1&9.51&14\\
RXJ1600.6-2159&16 00 40.57&-22 00 32.2&G9&1.43&10.2&8.44&3\\
RXJ1600.7-2127&16 00 42.77&-21 27 38.0&K7&0.77&11.7&8.92&10\\
RXJ1601.1-2113&16 01 08.01&-21 13 18.5&M0&0.68&12.0&8.80&9\\
RXJ1601.9-2008&16 01 58.23&-20 08 12.2&G5&1.62&9.9&7.67&2\\
RXJ1602.0-2221&16 02 00.39&-22 21 23.7&M1&0.60&12.8&8.84&12\\
RXJ1602.8-2401B&16 02 51.24&-24 01 57.4&K4&0.95&11.1&8.93&4\\
RXJ1602.8-2401A&16 02 52.43&-24 02 22.7&K0&1.35&10.4&7.65&1\\
RXJ1603.6-2245&16 03 35.50&-22 45 56.1&G9&1.43&10.6&8.36&3\\
RXJ1603.9-2031A&16 03 57.68&-20 31 05.5&K5&0.87&12.0&8.37&10\\
RXJ1604.3-2130&16 04 21.66&-21 30 28.4&K2&1.12&11.8&8.51&11\\
RXJ1605.6-2152&16 05 39.36&-21 52 33.8&M1&0.60&13.6&9.47&17\\
RXJ1606.2-2036&16 06 12.54&-20 36 47.3&K5&0.87&12.5&8.90&12\\
RXJ1607.0-2043&16 07 03.73&-20 43 07.4&M1&0.60&13.6&9.53&17\\
RXJ1607.0-2036&16 07 03.56&-20 36 26.5&M0&0.68&11.3&8.10&...\\
USco-155655.5-225839&15 56 55.46&-22 58 40.4&M0&0.68&13.2&9.43&14,15\\
USco-160142.6-222923&16 01 42.55&-22 29 23.9&M0&0.68&13.8&10.22&19\\
USco-160341.8-200557&16 03 41.87&-20 05 57.8&M2&0.49&13.7&9.49&18\\
USco-160343.3-201531&16 03 43.35&-20 15 31.5&M2&0.49&13.7&9.72&18\\
Usco-160428.4-190441&16 04 28.39&-19 04 41.4&M3&0.36&13.6&9.28&...\\
USco-160517.9-202420&16 05 17.92&-20 24 19.5&M3&0.36&13.3&9.14&15\\
USco-160643.8-190805&16 06 43.86&-19 08 05.6&K6&0.82&12.8&9.20&...\\
USco-160707.7-192715&16 07 07.67&-19 27 16.1&M2&0.49&13.8&9.80&19\\
USco-160801.4-202741&16 08 01.42&-20 27 41.7&K8&0.68&13.0&9.29&16\\
USco-160822.4-193004&16 08 22.34&-19 30 05.2&M1&0.60&12.9&9.06&12\\
USco-160823.2-193001&16 08 23.25&-19 30 00.9&K9&0.68&13.2&9.47&15\\
USco-160823.8-193551&16 08 23.88&-19 35 51.8&M1&0.60&13.3&9.25&...\\
USco-160825.1-201224&16 08 25.11&-20 12 24.6&M1&0.60&13.9&9.87&20\\
USco-160900.7-190852&16 09 00.76&-19 08 52.6&K9&0.68&13.1&9.15&15\\
USco-160908.4-200928&16 09 08.45&-20 09 27.8&M4&0.24&13.8&9.52&...\\
USco-160916.8-183522&16 09 16.85&-18 35 22.6&M2&0.49&14.0&9.67&20\\
USco-160946.4-193735&16 09 46.44&-19 37 36.1&M1&0.60&13.8&9.63&19\\
USco-160954.4-190654&16 09 54.41&-19 06 55.1&M1&0.60&13.7&9.60&18\\
USco-161031.9-191305&16 10 31.96&-19 13 06.2&K7&0.77&13.0&8.99&12\\
USco-161115.3-175721&16 11 15.34&-17 57 21.4&M1&0.6&13.2&9.20&15\\
USco-161347.5-183459&16 13 47.51&-18 35 00.4&M2&0.49&14.1&9.91&20\\
USco-161358.1-184828&16 13 58.15&-18 48 29.0&M2&0.49&14.0&9.88&20\\
GSC 06764-01305&15 35 57.80&-23 24 04.6&K3&0.99&12.0&9.43&11\\
GSC 06195-00768&15 57 02.34&-19 50 42.0&K7&0.77&11.1&8.37&...\\
GSC 06191-00019&15 59 02.09&-18 44 14.3&K6&0.82&11.1&8.11&...\\
GSC 06191-00552&15 58 47.70&-17 57 59.0&K3&0.99&11.5&8.33&5\\
GSC 06204-00812&16 03 02.69&-18 06 05.0&K4&0.95&11.3&8.73&5\\
GSC 06204-01067&16 03 23.68&-17 51 42.3&M2&0.49&12.4&8.61&...\\
GSC 06208-00834&16 06 31.70&-20 36 23.3&K6&0.82&12.4&8.73&10\\
GSC 06209-00735&16 08 14.74&-19 08 32.8&K2&1.12&11.0&8.43&5\\
GSC 06205-00954&16 08 31.38&-18 02 41.4&M0&0.68&12.2&8.91&9\\
GSC 06209-01501&16 08 56.73&-20 33 46.0&K5&0.87&11.9&8.62&9\\
GSC 06213-01358&16 09 30.30&-21 04 58.9&M0&0.68&12.1&8.92&9\\
GSC 06213-00194&16 09 40.99&-22 17 59.4&M0&0.68&11.6&8.44&7\\
GSC 06213-00306&16 10 42.03&-21 01 32.0&K5&0.87&11.9&8.56&6\\
GSC 06793-00868&16 11 56.33&-23 04 05.1&M1&0.6&12.2&8.82&...\\
GSC 06793-00797&16 13 02.72&-22 57 44.6&K4&0.95&11.7&8.46&8\\
GSC 06213-00306&16 13 18.59&-22 12 48.9&G9&1.43&9.8&7.43&1,2\\
GSC 06793-00994&16 14 02.12&-23 01 02.2&G4&1.63&10.9&8.61&4\\
GSC 06793-00806&16 15 34.57&-22 42 42.1&M1&0.60&11.2&7.91&7\\
GSC 06793-01406&16 16 17.95&-23 39 47.7&G7&1.56&9.9&8.10&2\\
GSC 06214-02384&16 19 33.96&-22 28 29.4&K0&1.35&10.5&8.51&3\\
GSC 06794-00480&16 20 45.96&-23 48 20.9&K3&0.99&11.9&8.93&8\\
GSC 06214-00210&16 21 54.67&-20 43 09.1&M1&0.60&11.6&9.15&8\\
GSC 06794-00537&16 23 07.83&-23 00 59.7&K2&1.12&11.0&8.18&4\\
GSC 06794-00156&16 24 51.36&-22 39 32.5&G6&1.59&9.3&7.08&1\\
GSC 06794-00337&16 27 39.56&-22 45 23.0&K1&1.25&10.9&8.08&6\\
GSC 06228-01359&16 35 48.36&-21 48 39.7&M0&0.68&12.4&8.48&10\\
ScoPMS015&15 57 19.99&-23 38 50.0&M0&0.68&12.4&8.88&...\\
ScoPMS017&15 57 34.31&-23 21 12.3&M1&0.60&12.9&8.99&14\\
ScoPMS019&15 59 59.95&-22 20 36.8&M1&0.60&12.3&8.63&11\\
ScoPMS021&16 01 25.64&-22 40 40.3&K1&1.25&13.6&8.52&16\\
ScoPMS022&16 02 08.45&-22 54 58.9&M1&0.60&13.4&9.55&17\\
ScoPMS027&16 04 47.76&-19 30 23.1&K2&1.12&11.0&8.04&5\\
ScoPMS028&16 05 27.27&-19 38 46.6&M1&0.60&13.3&9.55&16\\
ScoPMS042b&16 10 21.74&-19 04 06.7&M3&0.36&13.8&9.62&19\\
ScoPMS044&16 11 08.91&-19 04 46.9&K2&1.12&11.3&7.69&7\\
ScoPMS045&16 11 20.58&-18 20 54.9&K5&0.87&11.4&8.56&6\\
ScoPMS048&16 11 59.28&-19 06 53.3&K0&1.35&11.1&8.09&7\\
ScoPMS060&16 17 31.39&-23 03 36.0&G0&1.71&9.7&7.97&2\\
ScoPMS214&16 29 48.70&-21 52 11.9&K0&1.35&10.5&7.76&4\\
V1149 Sco&15 58 36.90&-22 57 15.0&G7&1.56&10.2&7.05&3\\
\enddata
\tablecomments{Typical uncertainties are $\sim$1 subclass for 
spectral types, $\sim$0.2 mag for $R$ magnitudes (taken from NOMAD; 
Zacharias et al. 2004), and $\sim$0.02 mag for $K$ magnitudes (taken 
from 2MASS; Skrutskie et al. 2006). The uncertainties in mass are 
dominated by systematic errors, including a global zero-point 
uncertainty of $\sim$20\% and individual uncertainties of as much as 
$\sim$100\% due to the possibility of further unresolved 
multiplicity.}
\end{deluxetable*}

\begin{deluxetable*}{lrrllrrrrrl}
\tabletypesize{\tiny}
\tablewidth{0pt}
\tablecaption{Known Binary Systems}
\tablehead{\colhead{Primary} & \colhead{RA} & \colhead{DEC} & 
\colhead{SpT$_{prim}$} & \colhead{$M_{prim}$} & \colhead{$R$} & 
\colhead{$K$} & \colhead{$\Delta$$K$} & \colhead{Sep} & \colhead{PA} 
& \colhead{Ref}
\\
\colhead{} & \multicolumn{2}{c}{(J2000)} & \colhead{} & 
\colhead{($M_{\sun}$)} & \colhead{(mag)} & \colhead{(mag)} &
\colhead{(mag)} & \colhead{(mas)} & \colhead{(deg)}
}
\startdata
GSC 06780-01061&16 06 54.36&-24 16 10.8&M3&0.36&12.2&8.86&1.3&1500&270.0&Kraus \& Hillenbrand (2007a)\\
GSC 06793-00569&16 13 29.29&-23 11 07.5&K1&1.25&11.1&8.49&2.7&1430&91.4&Metchev (2005)\\
GSC 06793-00819&16 14 11.08&-23 05 36.2&K0&1.35&10.0&7.46&0.21&222&304.8&Metchev (2005)\\
RXJ1600.5-2027&16 00 31.35&-20 27 05.0&M1&0.60&12.8&8.83&0.43&189&171.7&K\"ohler et al. (2000)\\
RXJ1601.7-2049&16 01 47.43&-20 49 45.8&M0&0.68&12.4&8.61&0.58&205&324.7&K\"ohler et al. (2000)\\
RXJ1601.8-2445&16 01 51.49&-24 45 24.9&K7&0.77&11.4&8.49&1.00&76&289.6&K\"ohler et al. (2000)\\
RXJ1602.9-2022&16 02 53.96&-20 22 48.1&K7&0.77&11.7&8.19&0.18&310&5.3&K\"ohler et al. (2000)\\
RXJ1603.9-2031B&16 03 54.96&-20 31 38.4&M0&0.68&12.5&8.62&0.53&121&140.9&K\"ohler et al. (2000)\\
RXJ1606.6-2108&16 06 37.41&-21 08 40.5&M1&0.60&13.2&9.11&0.09&1279&33.9&K\"ohler et al. (2000)\\
RXJ1607.0-1911&16 07 03.94&-19 11 33.9&M1&0.60&13.4&9.22&1.47&599&87.6&K\"ohler et al. (2000)\\
ScoPMS005&15 54 59.86&-23 47 18.2&G2&1.66&8.6&7.03&1.99&766&232.0&K\"ohler et al. (2000)\\
ScoPMS013&15 56 29.42&-23 48 19.8&M1.5&0.54&11.6&8.75&0.62&92&169.8&K\"ohler et al. (2000)\\
ScoPMS016&15 57 25.76&-23 54 22.0&M0.5&0.64&13.1&9.09&0.63&1324&226.0&K\"ohler et al. (2000)\\
ScoPMS020&16 01 05.19&-22 27 31.2&M3&0.36&12.9&8.75&0.60&193&313.7&K\"ohler et al. (2000)\\
ScoPMS023&16 02 10.45&-22 41 28.0&K5&0.87&10.2&8.06&0.65&300&345.6&K\"ohler et al. (2000)\\
ScoPMS029&16 05 42.67&-20 04 15.0&M2&0.49&13.4&9.16&0.56&643&352.6&K\"ohler et al. (2000)\\
ScoPMS031&16 06 21.96&-19 28 44.6&M0.5&0.64&12.8&8.62&0.64&578&148.2&K\"ohler et al. (2000)\\
ScoPMS042a&16 10 28.58&-19 04 47.0&M1&0.60&13.0&8.71&0.42&299&84.1&K\"ohler et al. (2000)\\
ScoPMS052&16 12 40.51&-18 59 28.3&K0&1.35&10.4&7.49&1.10&144&162.2&Metchev (2005)\\
\enddata
\tablecomments{Typical uncertainties are $\sim$1 subclass for 
spectral types, $\sim$0.2 mag for $R$ magnitudes (taken from NOMAD; 
Zacharias et al. 2004), and $\sim$0.02 mag for $K$ magnitudes (taken 
from 2MASS; Skrutskie et al. 2006). The uncertainties in masses are 
dominated by systematic errors, including a global zero-point 
uncertainty of $\sim$20\% and individual uncertainties of as much as 
$\sim$100\% due to the possibility of further unresolved 
multiplicity. Typical uncertainties in binary properties are 
$\sim$0.1 mag in $\Delta$$K$, $\sim$10 mas in separation, and 
$\sim$1 deg in PA.}
\end{deluxetable*}

\section{Observations and Data Analysis}

\subsection{Observations}

We observed our target sample in April-July 2007 with the Keck-II 10m and 
Palomar Hale 200" telescopes. All observations were obtained using the 
facility adaptive optics imagers, NIRC2 and PHARO. Both instruments have 
aperture masks permanently installed at or near the pupil plane in filter 
or pupil-stop wheels. The seeing quality was well above average for most 
of the Keck observations, yielding superb AO correction for bright targets 
and acceptable strehl ratios ($\sim$15-20\%) even for targets as faint as 
$R\sim$14. The Palomar observations were obtained under approximately 
median conditions ($\sim$1\arcsec\, seeing).

All observations conducted at Keck were obtained with a $K'$ filter, 
while those conducted at Palomar were obtained with a methane short 
filter, which is in $H$-band (central wavelength 1.57\,$\mu$m, 
bandpass 0.1\,$\mu$m). This filter was used instead of full $H$-band 
because of calibration errors related to dispersion that had been 
found in previous data sets. This strategy allowed us to achieve 
similar resolution limits at both telescopes, despite the smaller 
aperture size at Palomar. Our Palomar observations suffered a modest 
loss of sensitivity since the strehl is lower in $H$ than in $K'$, 
but the typical sensitivity limit in $H$ still allows us to detect a 
$\sim$30 $M_{Jup}$ companion at 40 mas for half of our sample 
members. Observing in $K$ would have yielded limits of 
$\Delta$$K$$\la$1 magnitude deeper (equivalent to 
$\Delta$$H$$\la$1.5 since low-mass companions are redder in $H-K$), 
and we decided this was not as important since the corresponding 
detection limits ($\ga$15-20 $M_{Jup}$) could not have reached the 
planetary mass range.

Observations at Keck used a 9 hole mask, with the longest baseline
8.27\,m and the shortest baseline 1.67\,m. We used a
multiple-correlated double sampling readout in a 512x512 subarray of
the ALADDIN detector, with 16 endpoint reads and a 10\,s exposure per
frame. Observations at Palomar with PHARO also used a 9 hole mask, with the longest 
baseline 3.94\,m and the shortest baseline 0.71\,m. To maximise the 
number of reads, we used either a 256x256 or 150x150 sub-array mode in 
one quadrant of the HAWAII detector, with a total of 16 or 28 reads 
respectively per array reset. Every read was saved to disk, so that in 
post-processing each file could be split into sub-frames. Splitting the 
data into more sub-frames minimises sensitivity to changing seeing or AO 
instabilities, and using less sub-frames minimises sensitivity to readout 
noise. We found that for the typical magnitudes of our targets, 
signal-to-noise was optimized by using read pairs separated by one read: 
giving 862\,ms exposure times for the 256x256 sub-array mode, and 430\,ms 
exposure times for the 150x150 sub-array mode.

A key requirement for obtaining good contrast limits is the 
contemporaneous observation of calibrator sources, ideally single stars 
which are nearby on the sky and similar in both optical and near-infrared 
brightness. A typical observing mode for isolated field stars is to obtain 
several sets of observations for a science target, interspersing visits to 
calibrator stars between each science observation. As a result, 
observations for a single science target might require as many as six 
target acquisitions (three calibrators, plus three visits to the source). 
However, all of our science targets are located in close proximity on the 
sky ($<$10$^o$) and they span a continuous range of brightness, so we were 
able to use the same calibrator star for multiple science targets and to 
intercalibrate between science targets. To this end, we divided our sample 
into 20 groups of $\sim$4 similar-brightness stars each, then observed 
each group contemporaneously. Specifically, we visited each group member 
three times, plus we obtained one visit for each of two independent field 
calibrators. This allowed us to typically observe four science targets 
with a total of 14 acquisitions, for an average of 3.5 acquisitions per 
target. The average total time per acquisition was $\sim$4 minutes, so our 
strategy required $\sim$15 minutes per target.

We summarize the composition of our target groups and list the 
independent calibrators in Table 3. We also include the observation date 
and the mean $R$ and $K$ magnitudes for each group. Some of our groups 
are bigger or smaller because our acquisition images showed that several 
intended targets were resolved binaries (Section 3.3). When this 
occurred, we removed the binary system from our sample; in the case of 
Groups 12-15, we found a large number of binaries, so we rearranged the 
group composition at the telescope and eliminated Group 13.

Finally, a large fraction of our sample has been observed previously 
with high-resolution imaging (Brandner et al. 1996; Metchev 2005; 
Bouy et al. 2006), so we knew a priori whether these stars had known 
companions. However, many of our targets have been observed only 
with speckle imaging (K\"ohler et al. 2000) or have not been 
observed with any high-resolution techniques. For these sources, we 
decided to obtain quick observations in direct-imaging mode in order 
to screen out obvious binaries. This also allowed us to test for 
companions at separations outside the nominal limit of aperture-mask 
interferometry (240 mas at Palomar and 320 mas at Keck).

In Table 4, we list all of the sources that were observed with direct 
imaging and summarize the observations. We observed all of these sources 
with NIRC2 or PHARO using the smallest pixel scale available (10 or 25 mas 
pix$^{-1}$, respectively) and a two-point diagonal dither pattern. Faint 
stars were observed with a $K'$ or $K_s$ filter, while bright stars that 
would have saturated the detector were observed with a $Br\gamma$ filter, 
which attenuates flux by a factor of $\sim$10 relative to broadband $K$ 
filters.

\begin{deluxetable*}{rllcccc}
\tabletypesize{\tiny}
\tablewidth{0pt}
\tablecaption{Aperture Mask Observations}
\tablehead{\colhead{Group} & \colhead{Science Targets} &
\colhead{Calibrator Stars} & \colhead{Telescope} & 
\colhead{$R$} & \colhead{$K$} &  \colhead{Epoch}
\\
\multicolumn{4}{c}{} & \colhead{(mag)} & \colhead{(mag)} & 
\colhead{(JD-2450000)}
}
\startdata
1&GSC 06213-00306, GSC 06794-00156&2M1618-2245, 2M1613-2218&Keck&9.3-10.4&7.08-8.02&4257\\
&RXJ1558.2-2328, RXJ1602.8-2401A\\
2&GSC 06213-00306, GSC 06793-01406&2M1602-1945, 2M1617-2320&Palomar&9.7-9.9&7.43-8.10&4251\\
&RXJ1601.9-2008, ScoPMS060\\
3&GSC 06214-02384, RXJ1600.6-2159&2M1559-2303, 2M1620-2231&Palomar&10.2-10.6&7.05-8.36&4252\\
&RXJ1603.6-2245, V1149 Sco\\
4&GSC 06793-00994, GSC 06794-00537&2M1613-2311, 2M1630-2118&Palomar&10.5-11.1&7.76-8.93&4312\\
&RXJ1602.8-2401B, ScoPMS214\\
5&GSC 061901-00552, GSC 06204-00812&2M1558-1747, 2M1606-1924&Palomar&11.0-11.5&8.04-8.73&4250\\
&GSC 06209-00735, ScoPMS027\\
6&GSC 06213-00306, GSC 06794-00337&2M1610-1818, 2M1629-2245&Palomar&10.9-11.9&8.08-8.56&4251\\
&ScoPMS045\\
7&GSC 06213-00194, GSC 06793-00806&2M1609-2216, 2M1611-1906&Palomar&11.1-11.6&7.69-8.44&4252\\
&ScoPMS044, ScoPMS048 A\\
8&GSC 06793-00797, GSC 06794-00480&2M1622-2036, 2M1558-2412&Palomar&11.4-11.9&8.46-9.15&4252\\
&GSC 06214-00210, RXJ1558.1-2405\\
9&GSC 06205-00954, GSC 06209-01501&2M1601-2123, 2M1608-2022&Palomar&11.9-12.2&8.62-8.92&4250\\
&GSC 06213-01358, RXJ1601.1-2113\\
10&GSC 06208-00834, GSC 06228-01359&2M1602-2133, 2M1635-2204&Palomar&11.7-12.4&8.37-8.92&4250\\
&RXJ1600.7-2127, RXJ1603.9-2031A\\
11&GSC 06764-01305, GSC 06793-00868&2M1613-2303, 2M1535-2330&Keck&11.8-12.3&8.51-9.43&4257\\
&RXJ1604.3-2130, ScoPMS019\\
12&RXJ1557.8-2305, RXJ1602.0-2221&2M1608-1916&Keck&12.4-13.0&8.61-9.27&4257\\
&RXJ1606.2-2036, USco-160822.4-193004\\
&USco-161031.9-191305\\
14&RXJ1550.0-2312, RXJ1600.2-2417&2M1600-2421, 2M1543-1929&Keck&12.8-13.0&8.99-9.29&4256\\
&ScoPMS017, USco-155655.5-225839\\
15&USco-155655.5-225839, USco-160517.9-202420&2M1557-2251, 2M1611-1802&Keck&13.1-13.3&9.14-9.47&4257\\
&USco-160823.2-193001, USco-160900.7-190852\\
&USco-161115.3-175721\\
16&ScoPMS021, ScoPMS028&2M1606-1949, 2M1607-2027&Keck&13.3-13.6&8.52-9.55&4257\\
&USco-160801.4-202741, USco-160823.8-193551&2M1601-2227\\
17&RXJ1551.1-2402, RXJ1605.6-2152&2M1550-2412, 2M1607-2050&Keck&13.4-13.6&9.47-9.73&4256\\
&RXJ1607.0-2043, ScoPMS022\\
18&USco-160341.8-200557, USco-160343.3-201531&2M1610-1904, 2M1614-1846&Keck&13.7-13.8&9.49-9.72&4256\\
&USco-160908.4-200928, USco-160954.4-190654\\
19&ScoPMS042b, USco-160142.6-222923&2M1602-2229, 2M1607-1929&Keck&13.8-13.8&9.62-10.22&4256\\
&USco-160707.7-192715, USco-160946.4-193735\\
20&USco-160825.1-201224, USco-160916.8-183522&2M1614-1846, 2M1608-2008&Keck&13.9-14.1&9.67-9.91&4256\\
&USco-161347.5-183459, USco-161358.1-184828\\
\enddata
\end{deluxetable*}

\begin{deluxetable}{lclcc}
\tabletypesize{\scriptsize}
\tablewidth{0pt}
\tablecaption{Direct Imaging Observations}
\tablehead{\colhead{Name} & \colhead{Telescope} & \colhead{$T_{int}$} & 
\colhead{Filter} & \colhead{Epoch}
\\
\colhead{} & \colhead{} & \colhead(s) & \colhead{} & 
\colhead{(JD-2450000)}
}
\startdata
GSC 06191-00019&Pal&56.64&Ks&4251\\
GSC 06195-00768&Pal&18.41&Ks&4199\\
GSC 06204-01067&Pal&56.64&Ks&4252\\
GSC 06205-00954&Pal&56.64&Ks&4250\\
GSC 06208-00834&Pal&56.64&Ks&4250\\
GSC 06209-01501&Pal&56.64&Ks&4250\\
GSC 06213-00194&Pal&56.64&Ks&4251\\
GSC 06213-00306&Pal&56.64&Ks&4251\\
GSC 06213-01358&Pal&56.64&Ks&4250\\
GSC 06214-00210&Pal&56.64&Ks&4251\\
GSC 06214-02384&Pal&56.64&Ks&4251\\
GSC 06764-01305&Pal&56.64&Ks&4252\\
GSC 06793-00797&Pal&56.64&Ks&4251\\
GSC 06793-00806&Pal&56.64&Ks&4251\\
GSC 06793-00868&Pal&56.64&Ks&4252\\
GSC 06793-00994&Pal&56.64&Ks&4251\\
GSC 06794-00156&Pal&56.64&Ks&4251\\
GSC 06794-00480&Pal&56.64&Ks&4251\\
GSC 06794-00537&Pal&56.64&Ks&4251\\
RXJ1550.0-2312&Keck&32&Brg&4256\\
RXJ1550.0-2312&Keck&32&Brg&4257\\
RXJ1550.9-2534&Keck&32&Brg&4257\\
RXJ1551.1-2402&Keck&32&Brg&4256\\
RXJ1557.8-2305&Keck&32&Brg&4257\\
RXJ1558.1-2405&Pal&56.64&Ks&4252\\
RXJ1558.2-2328&Pal&56.64&Ks&4251\\
RXJ1600.7-2127&Pal&56.64&Ks&4250\\
RXJ1601.1-2113&Pal&56.64&Ks&4250\\
RXJ1601.9-2008&Pal&56.64&Ks&4251\\
RXJ1602.0-2221&Keck&32&Brg&4257\\
RXJ1602.8-2401A&Keck&16&Brg&4257\\
RXJ1602.8-2401B&Pal&56.64&Ks&4251\\
RXJ1603.6-2245&Pal&56.64&Ks&4251\\
RXJ1603.9-2031A&Pal&56.64&Ks&4250\\
RXJ1604.3-2130&Pal&56.64&Ks&4252\\
RXJ1606.2-2036&Pal&56.64&Ks&4252\\
RXJ1607.0-2036&Pal&56.64&Ks&4251\\
ScoPMS015&Pal&56.64&Ks&4250\\
ScoPMS017&Keck&32&Brg&4256\\
ScoPMS019&Pal&56.64&Ks&4252\\
ScoPMS022&Keck&32&Brg&4256\\
ScoPMS027&Pal&28.32&Ks&4250\\
ScoPMS028&Keck&32&Brg&4257\\
ScoPMS042b&Keck&44&Brg&4256\\
ScoPMS044&Pal&56.64&Ks&4251\\
ScoPMS045&Pal&56.64&Ks&4251\\
ScoPMS048&Pal&56.64&Ks&4251\\
USco-160341.8-200557&Keck&32&Brg&4256\\
Usco-160428.4-190441&Keck&32&Brg&4257\\
USco-160517.9-202420&Keck&32&Brg&4257\\
USco-160643.8-190805&Pal&56.64&Ks&4252\\
USco-160707.7-192715&Keck&36&Brg&4256\\
USco-160801.4-202741&Pal&56.64&Ks&4252\\
Usco-160823.2-193001&Keck&32&Brg&4257\\
USco-160823.8-193551&Keck&32&Brg&4257\\
USco-160825.1-201224&Keck&32&Brg&4256\\
USco-160900.7-190852&Pal&56.64&Ks&4252\\
USco-160908.4-200928&Keck&32&Brg&4256\\
USco-160916.8-183522&Keck&32&Brg&4256\\
USco-160954.4-190654&Keck&32&Brg&4256\\
USco-161031.9-191305&Pal&56.64&Ks&4252\\
USco-161115.3-175721&Keck&32&Brg&4257\\
USco-161347.5-183459&Keck&32&Brg&4256\\
\enddata
\end{deluxetable}

\begin{figure}
\epsscale{1.00}
\plotone{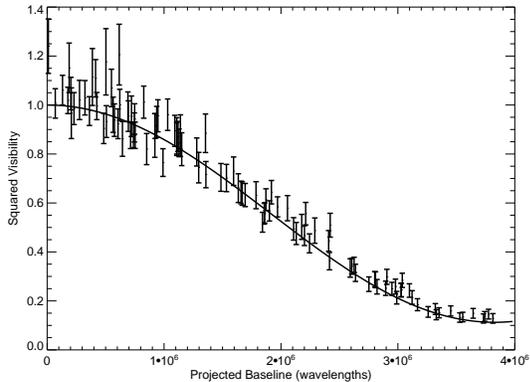}
\caption{The squared visibilities as a function of projected baseline for 
the 27 mas binary RXJ1550.0-2312. Despite a separation of only 0.6 $\lambda$$/D$, 
the binary system is clearly detected; the solid line denotes our 
best-fit value for the system parameters (Table 5).}
\end{figure}

\begin{figure}
\epsscale{1.00}
\plotone{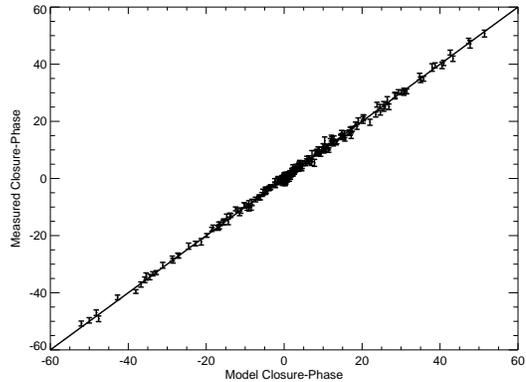}
\caption{
The measured close phases as a function of modeled closure phases for 
RXJ1550.0-2312, assuming that it has the best-fit parameters that we list 
in Table 5 (a 27 mas binary with a flux ratio of 2:1).}
\end{figure}

\subsection{Aperture Mask Analysis and Detection Limits}

The aperture-masking analysis pipeline is similar to that used for 
several previous papers containing Palomar masking data (Pravdo et al. 
2006; Lloyd et al. 2006; Martinache et al 2007). After subtracting 
the bias (dark) level, flat-fielding and removing bad pixels, the data are
windowed by a super-Gaussian (a function of the form $\exp(-k x^4)$). 
This window both limits sensitivity to readout noise and acts as a 
spatial-filter. Each frame is then Fourier-transformed, and the complex 
visibility extracted for each baseline. Complex visibilities cannot be
directly used for high-fidelity measurements because of their
sensitivity to variable optical 
aberrations and non common path errors. Instead, we use the averaged
squared visibility and the complex triple product
\citep{Lohmann83}. For each visit to each   
star, we extract squared-visibility, closure-phase, and the uncertainties 
on these quantities based on the scatter within one visit. Finally, the 
calibration process consists of estimating the instrumental 
squared-visibilities and closure-phases. The target star's squared 
visibilities are divided by the instrumental squared-visibilities and the 
instrumental closure-phase is subtracted from the measured closure-phase.

Figures 1 and 2 show a fit to data for the 27\,mas separation binary 
RXJ1550.0-2312. Squared visibility is plotted against baseline projected along the 
axis of the binary. As closure-phase is a multi-dimensional quantity, we 
chose to simply plot the measured closure-phases versus the model 
closure-phases. Despite this binary being at a separation of only 
0.6$\lambda/D$, it is clear that it is an extremely-high signal-to-noise 
detection. Note that only closure-phase was used in the fit: calibration 
errors are evident in the squared visibility data with the few points that 
have squared visibility greater than 1.0.

The error analysis procedure used to calculate the covariance matrix of 
closure-phase for each target is given in detail in Appendix~A. For all 
targets, an attempt at fitting a binary solution was made, by first 
searching exhaustively in a grid in position angle and separation at 
high-contrast and then by a gradient descent search to find the $\chi^2$ 
minimum. Errors in binary parameters were calculated from the curvature 
of the $\chi^2$ surface at the $\chi^2$ minimum (i.e. the same method as 
most least-squares algorithms). Detections were retained if their 
contrast was greater than a 99.9\% confidence threshold.

In order to calculate a detection threshold, we simulated 10000 data sets 
with the identical $(u,v)$-sampling and error properties of each target. 
For each of these simulated data sets, we calculated the best fit 
contrast ratio for every value of separation and position angle in a 
large grid, and then tabulated the maximum contrast ratio (i.e. brightest 
fitted companion) within a series of annuli. Our 99.9\% upper limits to companion 
brightness within each annulus was taken to be the contrast ratio where 
99.9\% of the simulations had no fitted companion brighter than this 
limit anywhere within the annulus. Details of the simulation and fitting  
algorithms can be found in the Appendix~A.

\subsection{Imaging Analysis and Detection Limits}

The imaging data were flat-fielded and dark- and bias-subtracted using 
standard IRAF procedures. We measured PSF-fitting photometry and 
astrometry for our sources using the IRAF package DAOPHOT (Stetson 1987), 
and specifically with the ALLSTAR routine. Where possible, we analyzed 
each frame separately in order to estimate the uncertainty in individual 
measurements and to allow for the potential rejection of frames with 
inferior AO correction; our final results represent the mean value for all 
observations in a filter. If the companion could not be easily 
distinguished in a single filter, we measured photometry from the coadded 
sum of all images at each dither position.

In all cases, we used the science target (or the primary star of a 
binary) to produce an analytic PSF composed of a gaussian core with 
lorentzian wings. If the science target was a close binary, we 
reconstructed the PSF using the algorithm described in Kraus \& 
Hillenbrand (2007b), which iteratively fits a template PSF to the 
primary and then subtracts the secondary to fit an improved estimate 
of the primary. Three sources appeared to be marginally detected in 
our observations; we retained these sources in our aperture mask 
sample and later confirmed their multiplicity (Section 4), so we 
processed their images with our PSF reconstruction routine and 
report the results. Finally, we calibrated our photometry using the 
known 2MASS $K_s$ magnitudes for each of our science targets; these 
absolute magnitudes are uncertain by $\sim$0.1-0.2 magnitudes due to 
the intrinsic variability of young stars (resulting from accretion 
or rotation).

At small separations ($\la$5$\lambda$$/D$), our imaging data for 
apparently single stars will be superceded by our aperture mask data, so 
the detection limits are not important. At larger separations 
($\ga$5$\lambda$$/D$), where spurious sources corresponding to AO speckles 
dominate, we adopted the detection limits suggested by Metchev (2005) for 
similar observations: $\Delta$$K=4$ at 250-500 mas, $\Delta$$K=5$ at 
500-1000 mas, and the sky background limit ($K\sim$16.5-17.5) at 
separations of $\ga$1\arcsec. We tested these limits for a set of the 
apparently single stars in our sample by subtracting an analytic PSF from 
the science target, then compiling the statistics for all apparently 
spurious detections. In all cases, the AO speckles fall at least a 
magnitude below our adopted limits.

Finally, the NIRC2 images were distortion-corrected using new high-order 
distortion solutions (P.B. Cameron, in prep) that deliver a significant 
performance increase as compared to the solutions presented in the NIRC2 
pre-ship 
manual\footnote{http://www2.keck.hawaii.edu/realpublic/inst/nirc2/}; the 
typical absolute residuals for bright, well-resolved stars are $\la$1 mas 
in narrow camera mode. The PHARO images were distortion-corrected using 
the solution derived by Metchev (2005), with fractional uncertainties in 
relative astrometry of $\sim$0.15\%. These uncertainties limit our 
astrometry for most close, well-resolved binary systems. The uncertainty 
for wider ($\ga$2-3\arcsec) pairs seems to be driven by variation due to 
differential tilt jitter, while the uncertainty for close blended pairs is 
driven by our ability to accurately model the single-star PSF.

\subsection{Stellar and Companion Properties}

Stellar properties can be difficult to estimate, particularly for 
young stars, since pre-main-sequence stellar evolutionary models are 
not well-calibrated. The mass of a given sample could be 
systematically uncertain by as much as 20\% (e.g. Hillenbrand \& 
White 2004), and individual masses could be uncertain by factors of 
2 or more due to unresolved multiplicity or the intrinsic 
variability that young stars often display (from accretion or 
rotational modulation of star spots). This suggests that any 
prescription for determining stellar properties should be treated 
with caution.

We estimated the properties of all of our sample members using the 
methods described in Kraus \& Hillenbrand (2007a). This procedure 
combines the 5 Myr isochrone of Baraffe et al. (1998) and the 
temperature scales of Schmidt-Kaler (1982) and Luhman et al. (2003) 
to directly convert observed spectral types to masses. Relative 
properties (mass ratios $q$) for all binaries in our sample were 
calculated by combining these isochrones and temperature scales with 
the empirical NIR colors of Bessell \& Brett (1998) and the K-band 
bolometric corrections of Leggett et al. (1998) to estimate $q$ from 
the observed flux ratio $\Delta$$K$. We also used these techniques 
to estimate masses for all of our sample members, which we list in 
Tables 1 and 2.

For all binary systems, we have adopted the previously-measured 
(unresolved) spectral type for the brightest component and 
inferred its properties from that spectral type. This should be 
a robust assumption since equal-flux binary components will 
have similar spectral types and significantly fainter 
components would not have contributed significant flux to the 
original discovery spectrum. Projected spatial separations are 
calculated assuming the mean distance of Upper Sco, 145$\pm$2 
pc (de Zeeuw et al. (1999). If the total radial depth of Upper 
Sco is equal to its angular extent ($\pm$8$^o$ or $\pm$20 
pc), then the unknown depth of each system within Upper Sco 
implies an uncertainty in the projected spatial separation of 
$\pm$14\%. The systematic uncertainty due to the uncertainty in 
the mean distance of Upper Sco is negligible in comparison 
($\la$2\%).

Finally, the sensitivity limits for some of our sample members 
extend to the bottom of the brown dwarf mass range and could 
potentially encompass the top of the planetary mass range. However, 
mass estimates for young giant planets are completely uncalibrated 
and there are ongoing debates regarding their peak and typical 
luminosities. The models of Baraffe et al. (2003) imply that a 
survey sensitive to $K\sim$16 could detect 7-10 $M_{Jup}$ planets at 
the distance and age of Upper Sco. However, more detailed models of 
planet formation by Marley et al. (2007) suggest that the typical 
luminosity of a young planet could be 1-2 orders of magnitude lower 
than previously predicted. These models differ primarily in their 
treatment of the initial conditions; recent models suggest that 
accretion shocks could dispel much of the initial energy, leading to 
lower internal entropy and correspondingly lower initial 
temperatures than the earlier models predicted. We can not currently 
resolve this controversy, so we only note that our limits on the 
presence of massive planets should be considered with caution.

\begin{figure*}
\epsscale{1.00}
\plotone{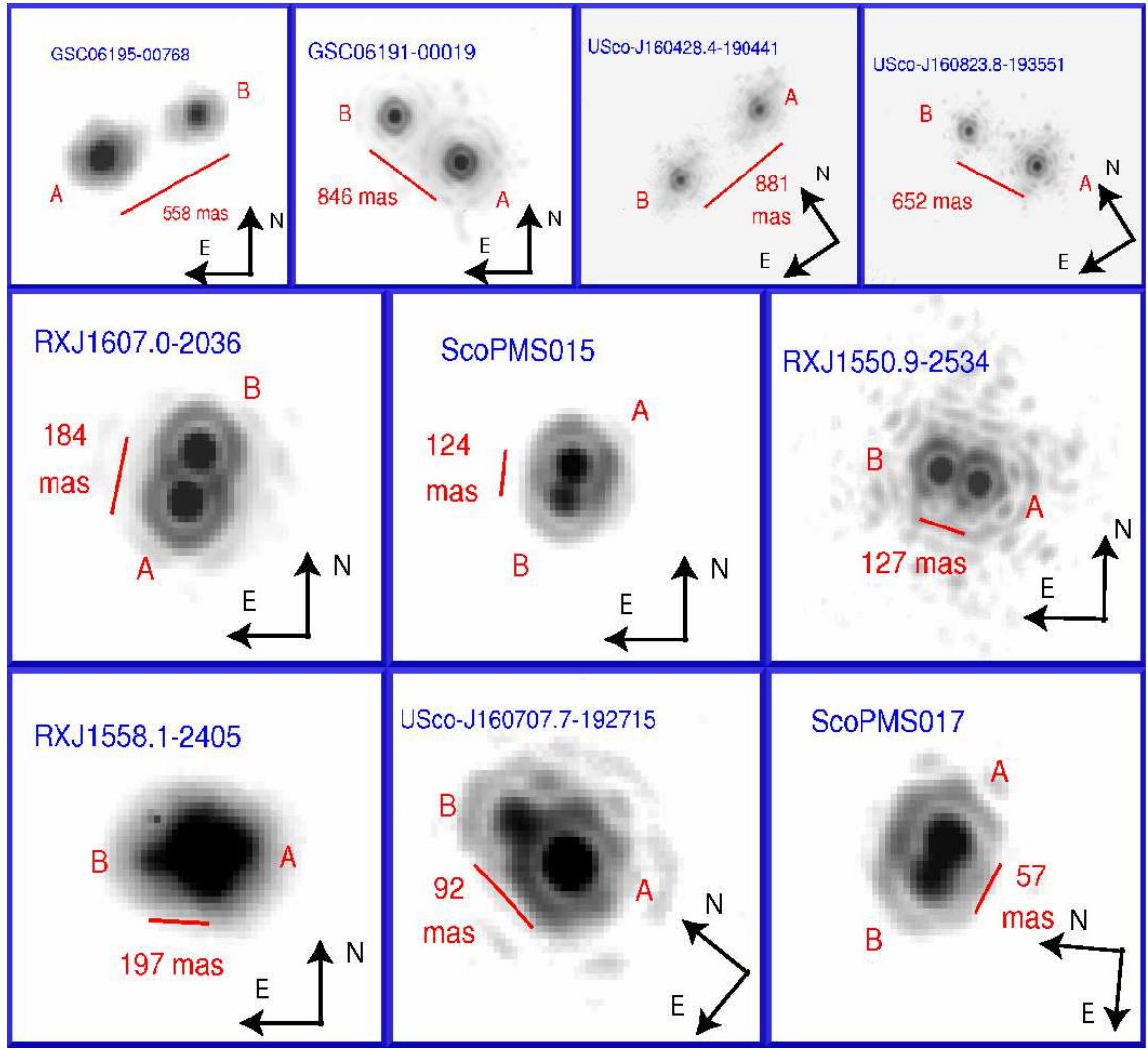}
\caption{Ten new systems that we observed with direct imaging. The top row 
shows relatively wide (0.5-1.0\arcsec) pairs, the middle row shows close, 
equal-flux pairs that are still easily distinguished, and the bottom row 
shows three very close or unequal-flux systems that could be difficult to 
identify with direct imaging alone, but were easily identified with 
aperture masking.}
\end{figure*}

\begin{figure*}
\epsscale{1.00}
\plotone{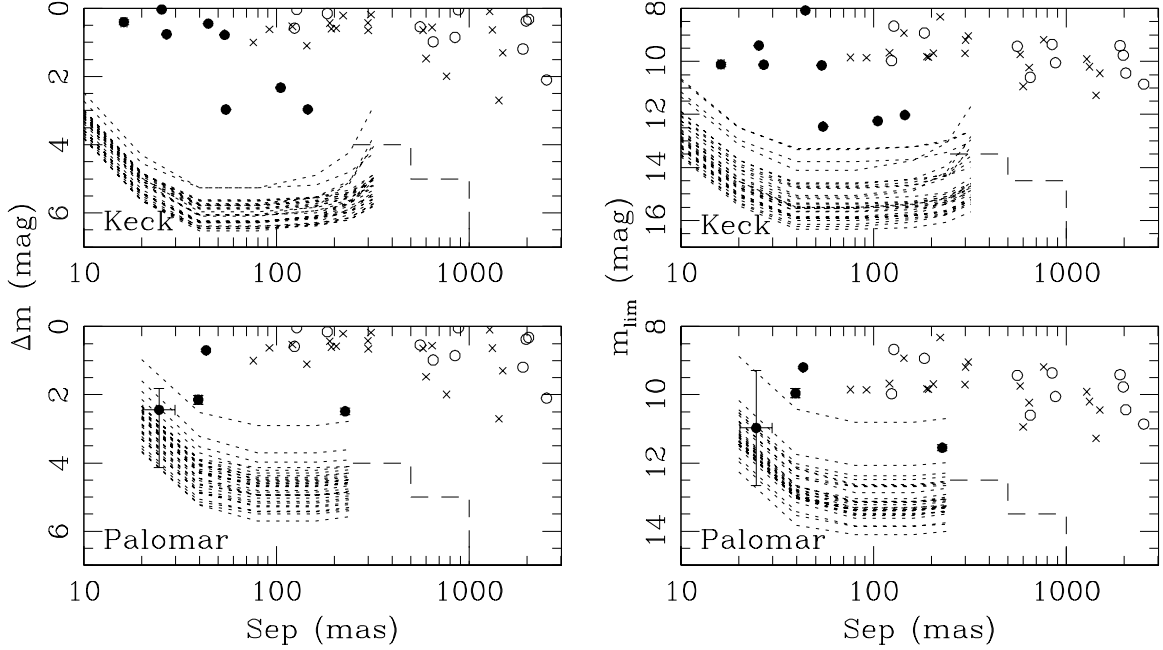}
\caption{Contrast ratio (left) and secondary brightness (right) as a 
function of separation for our new systems identified via masking (filled 
circles) and imaging (open circles), plus all known binary systems 
(crosses). We also show the corresponding aperture-masking detection 
limits for all apparently single stars in our survey (short-dashed lines) 
and our adopted sensitivity limits for our imaging data (long-dashed 
line).}
\end{figure*}

\begin{figure*}
\epsscale{1.00}
\plotone{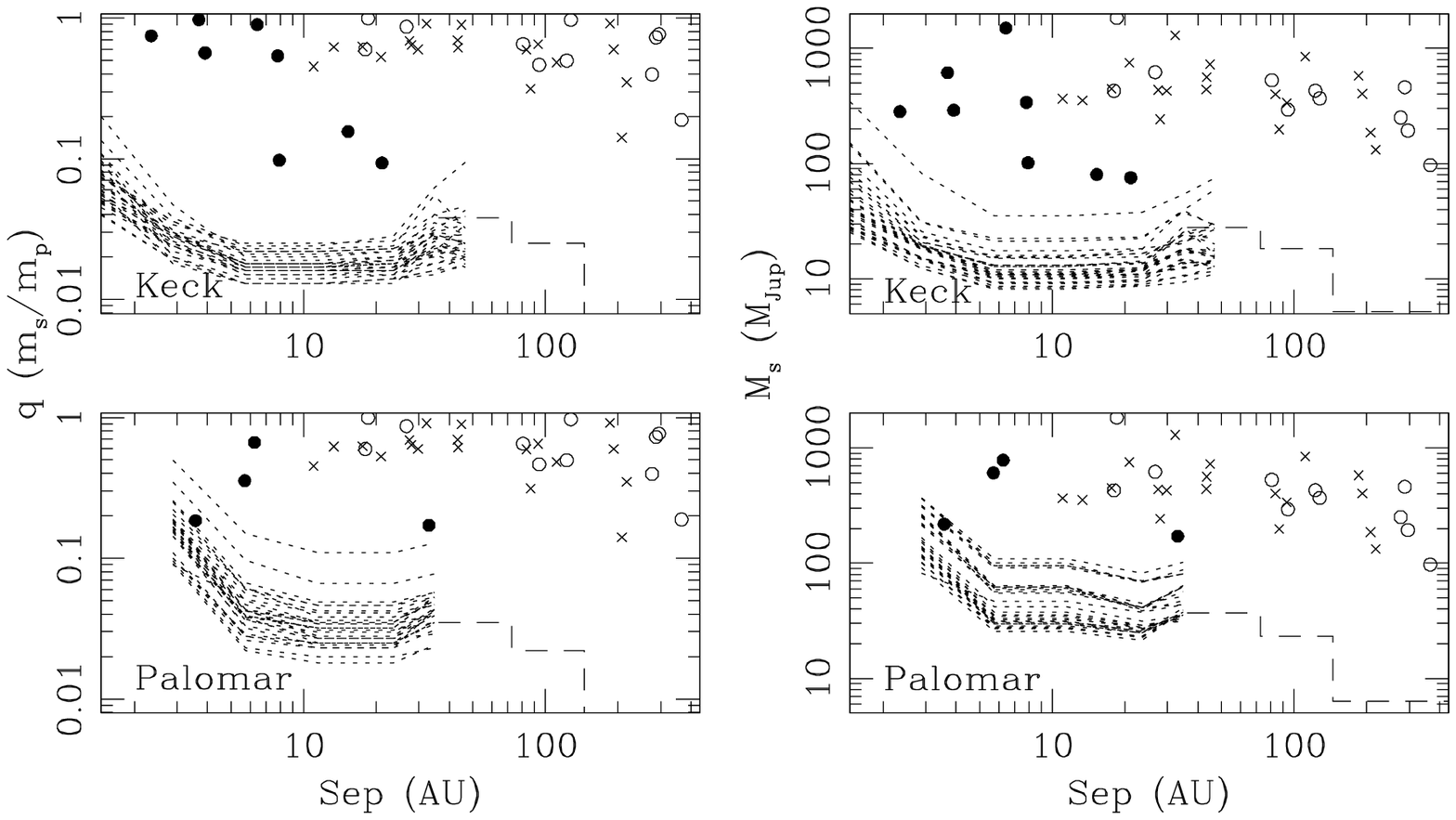}
\caption{As in Figure 4, but showing mass ratio (left) and secondary mass 
(right) as a function of separation.}
\end{figure*}

\section{New Companions in Upper Sco}

Our aperture mask survey is sensitive to companions with separations 
between $\sim$$\lambda$$/4D$ and $\sim$6$\lambda$$/D$ (at Keck) or 
$\sim$$4\lambda$$/D$ (at Palomar), corresponding to separation 
ranges of $\sim$10-320 mas and $\sim$20-240 mas, respectively. In 
this separation range, we identified 12 members of Upper Sco which 
possess a candidate companion at a confidence level of $\ga$99.5\% 
(99.9\% per annulus); the other 60 masking sample members appear to 
be single to within the detection limits we derived in Section 3.2. 
In Table 5, we list all of our newly-identified candidate companions 
and report their flux ratios, separations, and position angles. In 
Table 6, we summarize our derived upper limits as a function of 
separation for the 60 remaining members of our sample. The detection 
limits decline rapidly between $\sim$$\lambda$$/4D$ and 
$\sim$$2\lambda$$/3D$, but they are relatively flat at larger 
separations, extending to contrast ratios of $\sim$5.0-6.5 
magnitudes at Keck and $\sim$4.5-5.5 magnitudes at Palomar.

The system RXJ1550.0-2312 was observed on separate nights with separate 
calibrator sets at Keck in order to confirm the accuracy and 
repeatablity of our measurements. The separations and contrast 
ratios measured at each epoch agree to within $<$1 $\sigma$, 
suggesting that our results are repeatable and our assessed 
uncertainties are valid. We also we note that three of our 
calibrators appear to have companions. We did not use these 
observations in our final data calibration, and we report their 
astrometry in Table 4 for completeness. Finally, we note that 
the system GSC 06209-00735 has been previously identified as an 
SB1 by Guenther et al. (2007). The orbital period that they 
derived (2045$\pm$16 d) is consistent with the projected 
separation (25$\pm$5 mas; 3.6$\pm$0.7 AU) of our newly-imaged 
companion, so these detections appear to denote the same 
companion. Two more astrometric detections should allow us to 
resolve the orbital parameters which were not determined in the 
RV orbit ($K2$, $i$, and $\Omega$) and directly measure the 
masses of both stars.

In Table 7, we summarize the observed properties of 7 newly-detected 
binary systems discovered in our snapshot imaging program, three 
systems which were discovered in our aperture mask survey and 
subsequently recovered in our imaging data, and 9 known binary 
systems for which we report updated properties. In Figure 3, we show 
the corresponding discovery images for our newly-discovered 
binaries. We do not report any new candidate companions discovered 
outside a radius of 2\arcsec or with $K\ge$15 because of the 
significant probability that any such companions are background 
stars. We have previously estimated the density of background stars 
brighter than $K\sim$15 to be $\sim$1 arcmin$^{-2}$ (Kraus \& 
Hillenbrand 2007a), so the expected number of such contaminants 
inside $\sim$3\arcsec is only $\sim$1. However, there are many 
background bulge giants with apparent brightness $K\sim$16-17 that 
could be mistaken for faint wide companions, so an extension of 
these limits will require second-epoch observations to confirm 
common proper motion. Finally, we note that 4 of these sources fell 
near or inside the detection limits of the speckle interferometry 
survey of K\"ohler et al. (2000); their nondetection is most likely 
explained by an unfavorable orbital phase in 1994-1995 and 14 years 
of orbital motion.

In Table 8, we list the inferred stellar and binary properties 
for each of our newly-identified binary systems and the binary 
systems that we collated from the literature. In Figures 4 and 
5, we plot the contrast ratio and mass ratio versus the 
separation of all of our newly-detected companions, plus the 
corresponding detection limits for apparently single stars. The 
vast majority of our newly-identified candidate companions sit 
well above our survey's detection limits, suggesting that they 
are all valid detections. Their typical flux ratios 
($\Delta$$K\la$3) indicate that they have stellar masses. Both 
panels of Figure 3 show an obvious dearth of companions with 
flux ratios $\Delta$$K\ga$3, corresponding roughly to the 
substellar mass range. Our survey should have clearly detected 
any source in this range of parameter space, as has been proven 
for orbital monitoring of field binaries like GJ 802 (Lloyd et 
al. 2006; Ireland et al. 2008), so this deficit seems to 
represent a genuine absence of companions.

Finally, we did not detect any candidate companions near the typical 
detection limits of our survey, which correspond to 99.9\% confidence 
limits in any single separation bin or $\sim$99.5\% across all separation 
bins. We would expect an average of 0.3 false detections for the 60 
targets listed in Table 6, so our nondetection is consistent with the 
statistical estimate. We did detect four candidate companions with lower 
confidence levels (99.5-99.9\% in their separation bin, corresponding to 
overall confidence levels of $\sim$97.5\%-99.5\%). We would only expect to 
observe $\sim$1.5 false detections with this range of confidence levels, 
so 4 represents a marginally significant excess. A discussion off this
is given in Section 6.3.


\begin{deluxetable*}{lcrrrr}
\tabletypesize{\scriptsize}
\tablewidth{0pt}
\tablecaption{Companions Identified with the Aperture Mask}
\tablehead{\colhead{Primary} & \colhead{Telescope} & 
\colhead{$\Delta$$m$} & \colhead{Sep} & \colhead{PA}
\\
\colhead{} & \colhead{} & \colhead{(mag)} & \colhead{(mas)} & \colhead{(deg)}
}
\startdata
GSC 06209-00735\tablenotemark{a}&Palomar&2.44$\pm$1.16&24.6$\pm$5.2&42.5$\pm$3.6\\
GSC 06764-01305&Keck&2.97$\pm$0.01&54.68$\pm$0.16&173.76$\pm$0.19\\
GSC 06794-00156&Keck&0.45$\pm$0.01&44.30$\pm$0.07&230.74$\pm$0.08\\
RXJ1550.0-2312&Keck&0.76$\pm$0.01&26.95$\pm$0.05&222.13$\pm$0.13\\
RXJ1550.0-2312&Keck&0.76$\pm$0.01&26.93$\pm$0.04&222.07$\pm$0.11\\
RXJ1558.1-2405&Palomar&2.48$\pm$0.09&227.67$\pm$1.99&99.23$\pm$0.47\\
RXJ1601.9-2008&Palomar&2.14$\pm$0.13&39.31$\pm$1.57&217.67$\pm$0.59\\
ScoPMS017&Keck&0.78$\pm$0.01&53.86$\pm$0.19&68.93$\pm$0.20\\
ScoPMS019&Keck&0.03$\pm$0.01&25.40$\pm$0.12&113.55$\pm$0.62\\
ScoPMS027&Palomar&0.70$\pm$0.03&43.18$\pm$0.12&68.63$\pm$0.29\\
USco-160517.9-202420&Keck&0.40$\pm$0.07&16.15$\pm$0.59&251.12$\pm$1.11\\
USco-160707.7-192715&Keck&2.33$\pm$0.01&105.25$\pm$0.21&0.90$\pm$0.09\\
USco-161031.9-191305&Keck&2.96$\pm$0.02&145.55$\pm$0.43&81.63$\pm$0.14\\
Calibrators\\
2M1535-2330&Keck&1.35$\pm$0.01&92.35$\pm$0.17&311.46$\pm$0.09\\
2M1601-2227&Keck&0.64$\pm$0.09&249.76$\pm$0.5&328.73$\pm$0.14\\
2M1613-2218&Keck&3.97$\pm$0.07&93.33$\pm$1.04&11.9$\pm$0.6\\
\enddata
\tablenotetext{a}{The contrast ratio and separation are highly degenerate at separations 
this small, but at the least favorable separation, the secondary flux still represents a 
$>$7 $\sigma$ detection.}
\end{deluxetable*}

\begin{deluxetable*}{lcrrrrrrrrrrrr}
\tabletypesize{\tiny}
\tablewidth{0pt}
\tablecaption{Upper Limits for Undetected Companions}
\tablehead{\colhead{} & \colhead{} & 
\multicolumn{6}{c}{$\Delta$$m$\tablenotemark{a}} & \multicolumn{6}{c}{$q$ ($m_s/m_p$)\tablenotemark{a}}
\\
\colhead{Primary} & \colhead{Telescope} & 
\colhead{10-20} & \colhead{20-40} & \colhead{40-80} & 
\colhead{80-160} & \colhead{160-240} & \colhead{240-320} &  
\colhead{10-20} & \colhead{20-40} & \colhead{40-80} & 
\colhead{80-160} &\colhead{160-240} & \colhead{240-320} 
}
\startdata
GSC 06191-00552\tablenotemark{b}&Palomar&...&3.12&4.71&5.03&4.97&...&...&0.110&0.028&0.025&0.034&...\\
GSC 06204-00812&Palomar&...&3.23&4.81&5.13&5.03&...&...&0.100&0.029&0.025&0.041&...\\
GSC 06205-00954&Palomar&...&2.46&4.02&4.50&4.42&...&...&0.157&0.042&0.035&0.050&...\\
GSC 06208-00834&Palomar&...&2.71&4.30&4.81&4.66&...&...&0.161&0.036&0.030&0.043&...\\
GSC 06209-01501&Palomar&...&2.62&4.19&4.53&4.49&...&...&0.166&0.039&0.032&0.034&...\\
GSC 06213-00194&Palomar&...&2.34&3.93&4.43&4.30&...&...&0.185&0.043&0.036&0.048&...\\
GSC 06213-00306&Palomar&...&2.32&3.89&4.13&4.10&...&...&0.089&0.051&0.041&0.047&...\\
GSC 06213-00306&Keck&3.23&5.06&5.89&5.80&5.54&5.20&0.098&0.020&0.014&0.015&0.026&0.019\\
GSC 06213-00306&Palomar&...&3.69&5.24&5.71&5.57&...&...&0.092&0.022&0.018&0.024&...\\
GSC 06213-01358&Palomar&...&2.66&4.25&4.70&4.58&...&...&0.143&0.038&0.032&0.052&...\\
GSC 06214-00210&Palomar&...&1.90&3.52&3.98&3.91&...&...&0.250&0.060&0.046&0.053&...\\
GSC 06214-02384&Palomar&...&2.88&4.45&4.86&4.76&...&...&0.182&0.043&0.028&0.044&...\\
GSC 06228-01359&Palomar&...&2.70&4.29&4.65&4.57&...&...&0.143&0.038&0.032&0.052&...\\
GSC 06793-00797&Keck&3.57&5.40&6.25&6.10&5.75&4.94&0.056&0.019&0.013&0.014&0.024&0.023\\
GSC 06793-00797&Palomar&...&2.14&3.71&4.21&4.10&...&...&0.222&0.062&0.038&0.047&...\\
GSC 06793-00806&Palomar&...&0.97&2.52&2.90&2.78&...&...&0.499&0.150&0.109&0.128&...\\
GSC 06793-00994&Palomar&...&3.65&5.23&5.49&5.40&...&...&0.153&0.032&0.024&0.036&...\\
GSC 06793-01406&Palomar&...&3.06&4.61&4.95&4.82&...&...&0.187&0.056&0.042&0.053&...\\
GSC 06794-00337&Palomar&...&2.70&4.29&4.60&4.51&...&...&0.190&0.047&0.034&0.039&...\\
GSC 06794-00480&Palomar&...&1.59&3.23&3.70&3.60&...&...&0.349&0.097&0.066&0.077&...\\
GSC 06794-00537&Palomar&...&3.31&4.87&5.18&5.08&...&...&0.100&0.026&0.023&0.031&...\\
RXJ1551.1-2402&Keck&3.07&4.88&5.76&5.63&5.45&5.04&0.073&0.036&0.022&0.023&0.035&0.029\\
RXJ1557.8-2305&Keck&2.98&4.81&5.78&5.55&5.13&3.96&0.084&0.028&0.018&0.020&0.033&0.039\\
RXJ1558.2-2328&Keck&3.12&4.93&5.77&5.67&5.42&5.02&0.200&0.047&0.02&0.022&0.031&0.043\\
RXJ1600.2-2417&Keck&2.72&4.55&5.27&5.18&4.80&4.26&0.109&0.029&0.022&0.023&0.030&0.034\\
RXJ1600.6-2159&Palomar&...&2.93&4.52&4.95&4.86&...&...&0.177&0.043&0.027&0.042&...\\
RXJ1600.7-2127&Palomar&...&2.44&4.05&4.47&4.38&...&...&0.187&0.043&0.034&0.040&...\\
RXJ1601.1-2113&Palomar&...&2.64&4.23&4.68&4.62&...&...&0.143&0.039&0.032&0.053&...\\
RXJ1602.0-2221&Keck&2.48&4.30&5.26&4.90&4.23&2.86&0.136&0.038&0.024&0.028&0.062&0.094\\
RXJ1602.8-2401A&Keck&3.06&4.87&5.64&5.57&5.39&4.99&0.108&0.022&0.016&0.016&0.027&0.021\\
RXJ1602.8-2401B&Palomar&...&2.52&4.08&4.58&4.50&...&...&0.167&0.042&0.031&0.035&...\\
RXJ1603.6-2245&Palomar&...&3.10&4.65&4.94&4.90&...&...&0.143&0.037&0.027&0.029&...\\
RXJ1603.9-2031A\tablenotemark{b}&Palomar&...&2.86&4.45&4.94&4.86&...&...&0.143&0.033&0.027&0.039&...\\
RXJ1604.3-2130&Keck&3.57&5.43&6.23&6.15&5.79&5.50&0.060&0.018&0.013&0.013&0.021&0.017\\
RXJ1605.6-2152\tablenotemark{b}&Keck&3.20&5.05&6.09&5.93&5.66&5.24&0.064&0.030&0.017&0.018&0.029&0.025\\
RXJ1606.2-2036&Keck&2.99&4.83&5.72&5.54&5.05&4.05&0.094&0.025&0.017&0.018&0.022&0.034\\
RXJ1607.0-2043&Keck&3.15&4.99&5.85&5.78&5.56&5.16&0.066&0.030&0.019&0.020&0.028&0.025\\
ScoPMS021&Keck&3.37&5.19&6.06&5.94&5.75&5.05&0.081&0.020&0.013&0.014&0.016&0.020\\
ScoPMS022&Keck&3.14&4.97&5.96&5.84&5.63&5.21&0.066&0.030&0.018&0.019&0.029&0.025\\
ScoPMS028&Keck&3.02&4.85&5.64&5.54&5.40&4.86&0.078&0.031&0.021&0.022&0.027&0.029\\
ScoPMS042b&Keck&3.22&5.07&5.85&5.72&5.48&4.99&0.075&0.034&0.025&0.026&0.040&0.034\\
ScoPMS044&Palomar&...&2.48&4.05&4.38&4.29&...&...&0.182&0.05&0.035&0.053&...\\
ScoPMS045\tablenotemark{b}&Palomar&...&3.28&4.86&5.28&5.23&...&...&0.100&0.028&0.023&0.043&...\\
ScoPMS048&Palomar&...&2.34&3.93&4.29&4.20&...&...&0.258&0.067&0.049&0.057&...\\
ScoPMS060&Palomar&...&3.31&4.89&5.26&5.17&...&...&0.206&0.061&0.046&0.057&...\\
ScoPMS214&Palomar&...&3.58&5.14&5.41&5.30&...&...&0.096&0.023&0.020&0.023&...\\
USco-155655.5-225839&Keck&3.48&5.31&6.23&6.15&5.95&5.60&0.050&0.025&0.015&0.015&0.018&0.020\\
USco-160142.6-222923&Keck&3.18&5.00&5.79&5.68&5.52&5.00&0.067&0.027&0.018&0.019&0.023&0.025\\
USco-160341.8-200557&Keck&3.66&5.50&6.39&6.28&5.88&5.38&0.053&0.028&0.017&0.017&0.027&0.026\\
USco-160343.3-201531&Keck&3.82&5.65&6.52&6.34&6.05&5.66&0.049&0.024&0.016&0.017&0.021&0.023\\
USco-160801.4-202741&Keck&3.31&5.13&6.06&6.00&5.71&5.32&0.057&0.026&0.016&0.016&0.024&0.022\\
USco-160822.4-193004&Keck&3.30&5.13&6.05&5.82&5.08&3.76&0.060&0.029&0.017&0.019&0.040&0.046\\
USco-160823.2-193001&Keck&3.79&5.64&6.46&6.35&6.20&5.77&0.041&0.019&0.013&0.014&0.016&0.017\\
USco-160825.1-201224&Keck&3.64&5.46&6.28&6.09&5.91&5.43&0.048&0.024&0.016&0.017&0.021&0.023\\
USco-160900.7-190852&Keck&3.81&5.63&6.38&6.33&6.14&5.72&0.040&0.019&0.013&0.014&0.015&0.018\\
USco-160916.8-183522&Keck&3.57&5.40&6.25&6.14&5.97&5.44&0.055&0.029&0.018&0.018&0.020&0.025\\
USco-160946.4-193735&Keck&3.59&5.42&6.28&6.18&5.98&5.56&0.049&0.024&0.016&0.016&0.022&0.022\\
USco-160954.4-190654&Keck&3.59&5.42&6.26&6.09&5.68&4.96&0.049&0.024&0.016&0.017&0.030&0.028\\
USco-161115.3-175721&Keck&3.80&5.63&6.45&6.31&6.15&5.72&0.045&0.024&0.015&0.016&0.023&0.020\\
USco-161347.5-183459&Keck&2.99&4.83&5.61&5.53&5.33&4.88&0.077&0.036&0.023&0.024&0.055&0.032\\
USco-161358.1-184828&Keck&3.72&5.56&6.45&6.38&6.19&5.80&0.051&0.024&0.016&0.017&0.018&0.021\\
V1149 Sco&Palomar&...&3.49&5.06&5.43&5.35&...&...&0.154&0.038&0.025&0.035&...\\
\enddata
\tablenotetext{a}{The range of each separation bin is reported in 
units of mas, and the corresponding detection limits are reported 
in terms of $\delta$$m$ or $q$.}
\tablenotetext{b}{We detected candidate companions at lower 
confidence (97.5-99.5\%) for these four sources; we plan to obtain 
additional observations to confirm or disprove them.} 
\end{deluxetable*}

\begin{deluxetable*}{lcrrr}
\tabletypesize{\scriptsize}
\tablewidth{0pt}
\tablecaption{Companions Identified with Direct Imaging}
\tablehead{\colhead{Name} & \colhead{Telescope} & 
\colhead{$\Delta$$m$} & \colhead{Sep} &\colhead{PA}
\\
\colhead{} & \colhead{} & \colhead{(mag)} & \colhead{(mas)} & \colhead{(deg)}
}
\startdata
New\\
GSC 06191-00019&Palomar&0.85$\pm$0.01&845.8$\pm$1&58.0$\pm$0.1\\
GSC 06195-00768&Palomar&0.54$\pm$0.01&558$\pm$1&292.1$\pm$0.3\\
RXJ1550.9-2534&Keck&0.03$\pm$0.01&127.5$\pm$1&72.70$\pm$0.06\\
RXJ1558.1-2405\tablenotemark{a}&Palomar&1.86$\pm$0.03&197$\pm$2&98.8$\pm$0.3\\
RXJ1607.0-2036&Palomar&0.15$\pm$0.03&183.8$\pm$1&344.2$\pm$0.3\\
ScoPMS015&Palomar&0.58$\pm$0.02&124.1$\pm$1&166.5$\pm$0.4\\
ScoPMS017\tablenotemark{a}&Keck&0.65$\pm$0.01&57.1$\pm$1&68.34$\pm$0.11\\
USco-160428.4-190441&Keck&0.04$\pm$0.01&881.1$\pm$1&128.13$\pm$0.10\\
USco-160707.7-192715\tablenotemark{a}&Keck&1.59$\pm$0.01&91.8$\pm$1&2.1$\pm$0.3\\
USco-160823.8-193551&Keck&0.98$\pm$0.01&651.5$\pm$1&64.61$\pm$0.11\\
Known\\
GSC 06204-01067&Palomar&2.10$\pm$0.01&2528$\pm$4&93.04$\pm$0.02\\
GSC 06213-00306&Palomar&2.37$\pm$0.01&3186$\pm$5&305.11$\pm$0.01\\
GSC 06793-00806&Palomar&1.19$\pm$0.01&1907$\pm$3&338.81$\pm$0.03\\
GSC 06793-00868&Palomar&0.37$\pm$0.01&1981$\pm$4&155.29$\pm$0.06\\
RXJ1602.8-2401B&Palomar&2.91$\pm$0.02&7198$\pm$13&352.22$\pm$0.04\\
ScoPMS048&Palomar&1.76$\pm$0.01&3394$\pm$5&191.22$\pm$0.01\\
ScoPMS042b&Keck&2.48$\pm$0.03&4606$\pm$2&6.71$\pm$0.03\\
USco-160908.4-200928&Keck&0.32$\pm$0.01&2042$\pm$1&139.36$\pm$0.07\\
USco-161031.9-191305&Palomar&3.83$\pm$0.02&5775$\pm$9&112.66$\pm$0.02\\
\enddata
\tablenotetext{a}{Uncertainties are difficult to estimate due to to
  significant blending of the PSFs. The values and uncertainties from the
  aperture-masking detection in Table~5 should be used for this system.}
\end{deluxetable*}

\begin{deluxetable*}{lrllll}
\tabletypesize{\scriptsize}
\tablewidth{0pt}
\tablecaption{Companion Properties}
\tablehead{\colhead{Name} & \colhead{Sep} &
\colhead{$q$} & \colhead{$M_{prim}$} & \colhead{$M_{sec}$}
& \colhead{Source}
\\
\colhead{} & \colhead{(AU)} & \colhead{($m_s/m_p$)} & 
\colhead{($M_{\sun}$)} & \colhead{($M_{\sun}$)}
}
\startdata
GSC 06191-00019&122.6&0.50&0.82&0.41&Palomar-Imaging\\
GSC 06195-00768&80.9&0.65&0.77&0.50&Palomar-Imaging\\
GSC 06204-01067&366.6&0.19&0.49&0.09&Palomar-Imaging\\
GSC 06209-00735&3.6&0.18&1.12&0.21&Palomar-Masking\\
GSC 06213-00306&462.0&0.15&0.87&0.13&Palomar-Imaging\\
GSC 06214-00210&318.6&0.02&0.60&0.011&Palomar-Imaging\\
GSC 06764-01305&7.9&0.10&0.99&0.10&Keck-Masking\\
GSC 06780-01061&217.5&0.35&0.36&0.13&Kraus \& Hillenbrand (2007)\\
GSC 06793-00569&207.4&0.14&1.25&0.18&Metchev (2005)\\
GSC 06793-00806&276.5&0.40&0.60&0.24&Palomar-Imaging\\
GSC 06793-00819&32.2&0.91&1.35&1.23&Metchev (2005)\\
GSC 06793-00868&287.3&0.73&0.60&0.44&Palomar-Imaging\\
GSC 06794-00156&6.4&0.90&1.59&1.43&Keck-Masking\\
RXJ1550.0-2312&3.9&0.56&0.49&0.28&Keck-Masking (1)\\
RXJ1550.0-2312&3.9&0.56&0.49&0.28&Keck-Masking (2)\\
RXJ1550.9-2534&18.5&1.00&1.75&1.74&Keck-Imaging\\
RXJ1558.1-2405&33.0&0.17&0.95&0.16&Palomar-Masking\\
RXJ1558.1-2405&28.6&0.25&0.95&0.23&Palomar-Imaging\\
RXJ1600.5-2027&27.4&0.69&0.60&0.41&K\"ohler et al. (2000)\\
RXJ1601.7-2049&29.7&0.60&0.68&0.41&K\"ohler et al. (2000)\\
RXJ1601.8-2445&11.0&0.45&0.77&0.35&K\"ohler et al. (2000)\\
RXJ1601.9-2008&5.7&0.36&1.62&0.58&Palomar-Masking\\
RXJ1602.8-2401B&1043.7&0.10&0.95&0.10&Palomar-Imaging\\
RXJ1602.9-2022&45.0&0.90&0.77&0.69&K\"ohler et al. (2000)\\
RXJ1603.9-2031B&17.6&0.63&0.68&0.43&K\"ohler et al. (2000)\\
RXJ1606.6-2108&185.5&0.91&0.60&0.55&K\"ohler et al. (2000)\\
RXJ1607.0-1911&86.9&0.31&0.60&0.19&K\"ohler et al. (2000)\\
RXJ1607.0-2036&26.7&0.87&0.68&0.59&Palomar-Imaging\\
ScoPMS005&111.1&0.48&1.66&0.80&K\"ohler et al. (2000)\\
ScoPMS013&13.3&0.62&0.54&0.34&K\"ohler et al. (2000)\\
ScoPMS015&18.0&0.60&0.68&0.41&Palomar-Imaging\\
ScoPMS016&192.0&0.60&0.64&0.38&K\"ohler et al. (2000)\\
ScoPMS017&7.8&0.54&0.60&0.32&Keck-Masking\\
ScoPMS017&8.3&0.59&0.60&0.35&Keck-Imaging\\
ScoPMS019&3.7&0.97&0.60&0.58&Keck-Masking\\
ScoPMS020&28.0&0.64&0.36&0.23&K\"ohler et al. (2000)\\
ScoPMS023&43.5&0.61&0.87&0.53&K\"ohler et al. (2000)\\
ScoPMS027&6.3&0.66&1.12&0.74&Palomar-Masking\\
ScoPMS029&93.2&0.65&0.49&0.32&K\"ohler et al. (2000)\\
ScoPMS031&83.8&0.59&0.64&0.38&K\"ohler et al. (2000)\\
ScoPMS042a&43.4&0.70&0.60&0.42&K\"ohler et al. (2000)\\
ScoPMS048&492.1&0.30&1.35&0.40&Palomar-Imaging\\
ScoPMS052&20.9&0.53&1.35&0.71&Metchev (2005)\\
Usco-160428.4-190441&127.8&0.97&0.36&0.35&Keck-Imaging\\
USco-160517.9-202420&2.3&0.75&0.36&0.27&Keck-Masking\\
USco-160707.7-192715&15.3&0.16&0.49&0.08&Keck-Masking\\
USco-160707.7-192715&13.3&0.28&0.49&0.14&Keck-Imaging\\
USco-160823.8-193551&94.5&0.46&0.60&0.28&Keck-Imaging\\
USco-160908.4-200928&296.1&0.77&0.24&0.18&Keck-Imaging\\
USco-161031.9-191305&21.1&0.09&0.77&0.07&Keck-Masking\\
USco-161031.9-191305&837.4&0.04&0.77&0.03&Palomar-Imaging\\
\enddata
\tablecomments{Typical uncertainties in separations are $\sim$15\% 
and result from the unknown depth of each system within the 
association. The uncertainties in masses are dominated by systematic 
errors, including a global zero-point uncertainty of $\sim$20\% and 
individual uncertainties of as much as $\sim$100\% due to the 
possibility of further unresolved multiplicity. The mass ratio 
estimates should be more precise ($\sim$5-10\%) since many 
systematics (distance, age, extinction, and zero-point shifts) are 
cancelled, but they are still vulnerable to large systematic 
errors due to unresolved multiplicity.} 
\end{deluxetable*}

\begin{figure*}
\epsscale{1.00}
\plotone{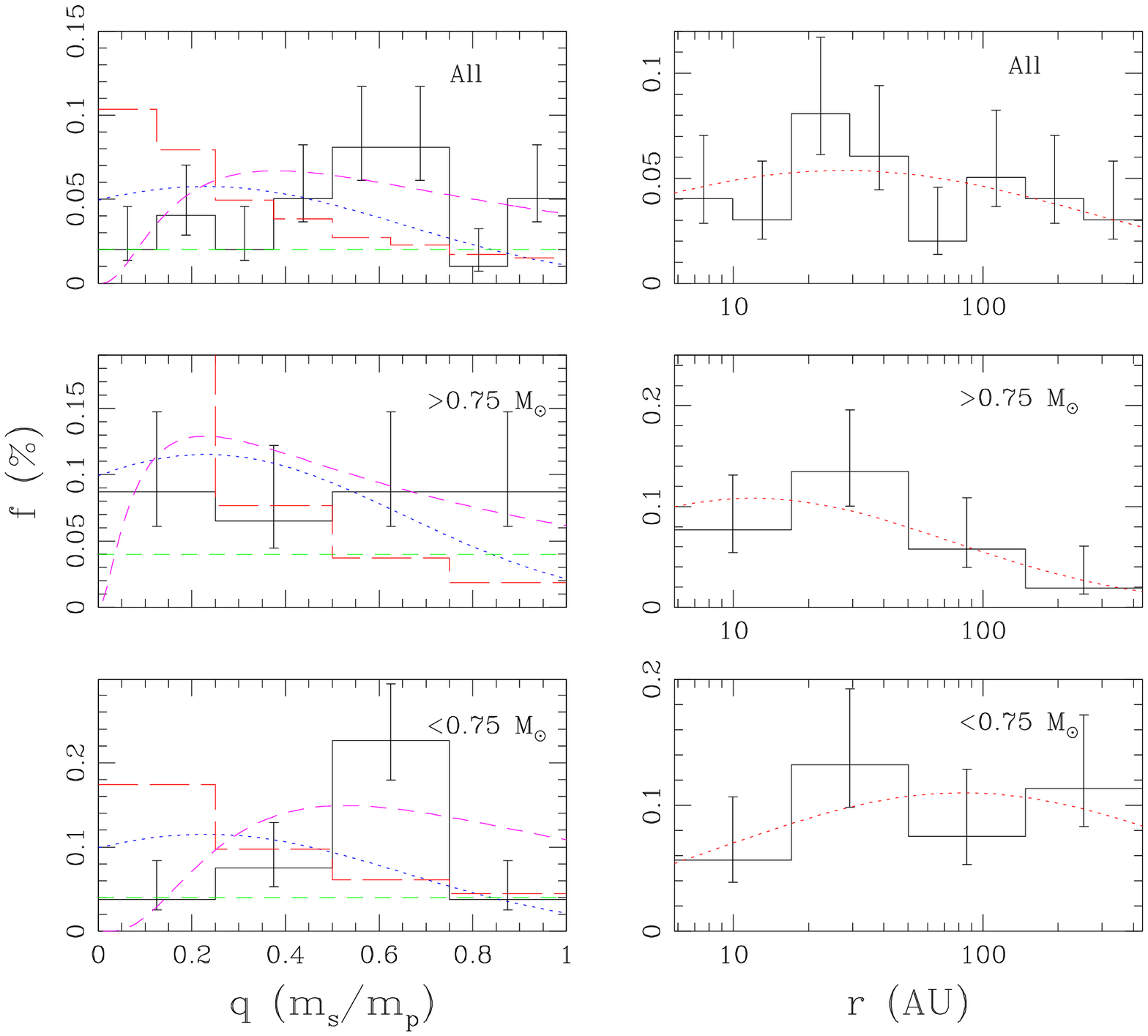}
\caption{The mass ratio distributions (left) and separation distributions 
(right) for all stars in our sample (top), the more massive half (FGK 
stars; $M>0.75$ $M_{\sun}$; middle), and the less massive half (M stars; 
$M<0.75$ $M_{\sun}$; bottom). On the left, we overplot several suggested 
mass functions: a truncated Gaussian distribution (blue dotted), a 
constant distribution (green dashed), a distribution of companions drawn 
from the IMF (red long-dashed), and the best-fit log-normal distribution 
(magenta dashed). On the right, we overplot the best-fit lognormal 
distribution (red dashed) for each subsample.}
\end{figure*}

\section{The Stellar Sea}

The properties of multiple star systems are important 
diagnostics for placing constraints on star formation 
processes. A comprehensive theory of star formation should be 
able to reproduce the observed separation distribution, mass 
ratio distribution, and total fraction of binary systems, as 
well as any mass or environmental dependences of these 
properties. The mass ratio distribution also plays a critical 
role in defining the brown dwarf desert since the bottom tail 
of the distribution represents the upper bound of the desert.

Most recent efforts to model binary formation have typically 
assumed that stellar and prestellar interactions play a key 
role in establishing binary properties. The most popular type 
of model assumes that a cluster of 5-10 protostellar embryos 
form from a single turbulently fragmenting cloud core (e.g. 
Kroupa 1995; Sterzik \& Durisen 1998; Kroupa \& Bouvier 2003; 
Kroupa et al. 2003; Delgado-Donate et al. 2003; Hubber et al. 
2005); these embryos would then undergo mass accretion and 
dynamical evolution to form single stars and stable multiple 
systems. However, other stellar properties place strong limits 
on the rate of early dynamical evolution. Close stellar 
encounters would tend to dissipate or truncate disks, with 
smaller stars having fewer and shorter-lived disks, but there 
is no evidence for this trend (e.g. White \& Basri 2003; Luhman 
2004; Scholz et al. 2006). Dynamical encounters might also 
eject lower-mass stars and brown dwarfs, but no such ejected 
population is seen (Luhman 2006), though some models suggest 
that strong ejections might be rare (Bate \& Bonnell 2005). 
Finally, any dynamically active environment would truncate the 
stellar binary separation distribution for all stars in the 
association. The absence of low-mass wide binaries has often 
been interpreted as a sign of this process, but this absence is 
seen even in environments where the wide binary frequency is 
very high for solar-mass stars (Kraus \& Hillenbrand 2007a), 
so it may have another explanation.

Modeling efforts that concentrate on other binary formation 
processes have not advanced sufficiently to make any rigorous 
prediction. These processes, which are summarized by Goodwin et 
al. (2007), include fragmentation of massive circumstellar 
discs, the role of magnetic support in prestellar cores, and 
fission of quasi-static rotating cores. All of these processes 
are more significant for isolated cores than for the 
dynamically-active turbulent fragmentation scenario discussed 
above, so the limits on dynamical activity of young stars 
suggest that they should be considered in more detail in the 
future.

Given the absence of theoretical predictions, we are left only 
with empirical comparisons to other samples. Previous field 
multiplicity surveys (DM91; Fischer \& Marcy 1992, hereafter 
FM92; Reid \& Gizis 1997, hereafter RG97) have suggested a 
range of possible results for the separation distribution, mass 
ratio distribution, and total frequency of binary systems. We 
will compare our results to these surveys and to the expected 
result if binary companions are drawn from an IMF. None of 
these explanations produce an acceptable fit for our mass ratio 
distribution, so the next step is to test other analytic 
distributions. Our number statistics do not support strong 
constraints on this analysis yet, so we will limit our analysis 
to a single functional form (the log-normal distribution) until 
we conclude the second half of our survey, an examination of 
young stars in Taurus. 

Finally, we note that two of the systems among our sample 
(USco-160428.4-190441 and USco-160825.1-201224) would have fallen 
below the optical flux limit of our sample ($R\le$14, imposed by the 
AO system) if the primaries were single. Including these systems in 
our statistical analysis would bias our results toward higher binary 
frequencies, so we have omitted them from our subsequent analysis. 
There is also an opposing effect due to the inclusion of unresolved 
binary pairs (which we identify as single stars) that would be 
omitted by the same criterion if we knew they were binaries. We 
obviously can't identify these systems, so we only note that the 
effect should be small. If the binary frequency at small separations 
is $\sim$10-20\% and the mass ratio distribution is similar to the 
distribution we observe, then no more than 1-2 unresolved systems 
should be included as ``single stars''. This systematic bias should 
not be significant compared to the statistical uncertainties in our 
results. We also note that our entire analysis must implicitly adopt 
the assumption that the mass ratio distribution and separation 
distribution are uncorrelated over the survey's separation range. 
This assumption has not been rigorously tested, but a simultaneous 
investigation of both parameters would require a far larger sample.

\begin{deluxetable}{lrllll}
\tabletypesize{\scriptsize}
\tablewidth{0pt}
\tablecaption{Binary Mass Ratio Distributions}
\tablehead{\colhead{Distribution} & \colhead{Masses} &  
\colhead{$\chi$$^2$} & \colhead{$P$} & \colhead{$D_{KS}$} & 
\colhead{$P$}
}
\startdata
Gaussian&All&21.2&0.007&0.35&0.00025\\
&High&5.1&0.28&0.34&0.047\\
&Low&13.6&0.009&0.41&0.0016\\
Constant (16\%)&All&31.9&0.0001&...&...\\
&High&11.5&0.021&...&...\\
&Low&18.3&0.0011&...&...\\
Constant (35\%)&All&11.4&0.18&0.19&0.14\\
&High&0.17&0.997&0.18&0.67\\
&Low&11.1&0.025&0.24&0.17\\
IMF&All&37.0&0.000005&0.46&0.0000004\\
&High&14.0&0.003&0.52&0.0003\\
&Low&21.3&0.00009&0.44&0.0005\\
Log-normal&All&9.7&0.14&0.15&0.41\\
&High&1.5&0.47&0.21&0.49\\
&Low&7.6&0.022&0.22&0.23\\
\enddata
\end{deluxetable}

\subsection{The Mass Ratio Distribution}

Observations of field stars have suggested that the mass ratio 
distribution is strongly dependent on mass. DM91 showed that F and G 
stars have mass ratio distributions biased toward inequal masses, 
roughly consistent with a truncated gaussian distribution (albeit 
with few constraints for $q<0.1$). By contrast, FM92 and RG97 found 
a distribution for early M dwarfs that is roughly flat, and numerous 
surveys have shown that the distribution for late-M dwarfs and brown 
dwarfs is biased toward equal masses ($q\ga0.7$; Close et al. 2003; 
Bouy et al. 2003; Burgasser et al. 2003). However, these surveys 
have all studied old field populations. Simulations show that 
dynamical evolution is typically not significant once a star enters 
the field (e.g. Weinberg et al. 1987), but a large fraction of stars 
are thought to be drawn from dense cluster environments (like the 
Orion Nebula Cluster or the Pleiades), so their properties could 
have been shaped by significant dynamical evolution in their natal 
environment. This suggests that primordial binary properties could 
differ significantly from those of their older brethren.

In the left side of Figure 6, we plot histograms of the mass 
ratio distribution for our entire sample of 99 stars, only the higher-mass 
stars (46 FGK dwarfs, representing masses $\ga$0.7 $M_{\sun}$), 
and only the lower-mass stars (55 M dwarfs, representing masses 
$\la$0.7 $M_{\sun}$). The mass ratio distributions are plotted 
for projected separations of 0.04-3.0\arcsec\, (6-435 AU), 
where the inner bound is defined by the inner limit for our 
survey to be sensitive to $q\sim$0.04 and the outer bound is 
defined by the field star contamination rate predicted for 
Upper Sco binaries by Kraus \& Hillenbrand (2007a). All of the 
number statistics are only moderately significant, but they 
still suffice for placing limits on the range of functional 
forms for the primordial mass ratio distribution.

In all three cases, our survey's mass ratio distribution is not strongly 
inconsistent with a constant distribution, so our ability to test more 
complex functional forms is limited. However, our data will suffice to 
test previously-suggested functions. To this end, we have compared our 
results to three distributions: a Gaussian distribution like that 
suggested by DM91, a constant distribution like that suggested by FM92, 
and a distribution which was assembled by assuming stars are randomly 
drawn from the IMF of Upper Sco. None of these functions feature a 
low-mass cutoff that could explain the brown dwarf desert, so we have also 
conducted preliminary tests of a new functional form: the log-normal 
distribution. Our constraints on this distribution are not very stringent, 
but they allow some preliminary conclusions. We will summarize the results 
for each of these tests in the following subsections, and we report the 
goodness of fit statistics (as measured with $\chi$$^2$ and 
Kolmogorov-Smirnov tests) in Table 9. The Kolmogorov-Smirnov test is 
more sensitive in cases where the bin size is a significant 
fraction of the total range of parameter space or when the trial
distribution function changes rapidly accross a bin, so all of our
conclusions are based on its results.

\subsubsection{The Gaussian Distribution}

DM91 found that the mass ratio distribution for field F and G 
dwarfs could be well-fit by a Gaussian distribution centered at 
low $q$ values ($\mu=0.23$, $\sigma=0.42$). Their survey was 
not sensitive to substellar companions ($q<0.1$), but if this 
functional form is valid, it suggests that substellar 
companions should be very common ($f\sim$10\%, with 4\% falling 
in our survey's separation range). However, there are no 
physical motivations for assuming that an arbitrarily chosen 
segment of a Gaussian function (-0.5$\sigma$ to +2.0$\sigma$) 
should predict the mass ratio distribution, so any similarity 
may be a coincidence. In the three left-hand panels of of 
Figure 6, we plot the $q$ distribution suggested by DM91 with a 
blue dotted line. This distribution was originally defined for all 
separations, but DM91 found that only 40\% of their systems fell within 
our survey's separation range, so we have scaled their function by this 
amount. This ensures that the overall binary frequency and the shape of 
the distribution are directly comparable.

Visual inspection shows that our full sample's $q$ distribution 
is more biased toward equal-mass companions than that of DM91, 
an observations that is supported by goodness-of-fit tests. 
This level of disagreement could be a result of our wider mass 
range than DM91's sample since lower-mass binary systems are 
thought to have mass ratios that are not as biased toward low 
masses. The relative levels of agreement for our high-mass and 
low-mass subsamples support this assessment; the high-mass 
subsample is only somewhat inconsistent with the DM91 
distribution, while the low-mass subsample is very 
significantly inconsistent.

\subsubsection{The Constant Distribution}

A field binary survey by FM92 found that the mass ratio 
distribution for field early M dwarfs seemed to be better fit 
by a flat distribution of mass ratios for $q\ga$0.4. RG97 found 
that this flat distribution extends to much lower mass ratio 
distributions for M dwarfs, though they also suggested the 
existence of a possible peak near unity ($q\ga$0.8). As with 
the Gaussian distribution, a flat distribution would suggest 
that substellar companions are not uncommon relative to stars, 
but these survey were not sensitive enough to actually detect 
most brown dwarf secondaries. Their total binary fractions 
(20$^{+7}$$_{-5}$\% for $q>0.4$ or 16$^{+7}$$_{-4}$\% for all 
$q$, respectively, in this separation range) are marginally 
inconsistent, but the RG97 sample (which is more rigorously 
volume-limited) contains 17 of the 37 binary 
systems considered by FM92, so we will adopt their value. In 
the three left-hand panels of Figure 6, we plot the flat $q$ 
distribution suggested by RG97 with a green dashed line.

Visual inspection suggests that a constant distribution might 
be more appropriate for our sample's $q$ distribution than the 
DM91 Gaussian distribution. However, the flat distribution of 
RG97 appears to fall systematically too low for the full sample 
and both subsamples, yielding high $\chi$$^2$ values. If we 
renormalize the flat distribution to match our overall binary 
frequency (36$^{+5}_{-4}\%$), we find much better agreement. 
The corresponding Kolmogorov-Smirnov tests, which only measure 
the cumulative density function and implicity include our 
renormalization, also find that a constant distribution is 
consistent or perhaps marginally inconsistent. We also note 
that we found no clear evidence of an excess of equal-mass 
binaries; the 2-sigma upper limit in the highest-mass 
bin of our entire sample ($q>0.875$) is $f<$11.4\%.

\subsubsection{A Distribution Drawn from the IMF}

Some theories also suggest that binary companions could be 
drawn randomly from the IMF. This idea used to be popular since 
it could be naturally explained as a result of random pairing 
and because previous results were moderately consistent, but it 
has fallen out of favor as the role of dynamical interactions 
has been increasingly constrained. However, an IMF could still 
be valid for wide binaries (which may form during the turbulent 
fragmentation of a large cloud core) and it is not clear where 
this regime ends and where the binary fragmentation of a 
collapsing protostar begins. This suggests that it would be 
prudent to test the validity of an IMF-based $q$ distribution. 
We adopted our IMF (hereafter the companion mass function, or 
CMF) from the spectroscopic membership surveys of Preibisch et 
al. (1998, 2002), and Slesnick et al. (2006a, 2007); this IMF 
can be described by a broken power law, 
$\Psi$$(M)=dN/dM$$\propto$$M^{-\alpha}$, where $\alpha=-2.8$ 
for $0.6<M<2.0$ $M_{\sun}$, $\alpha=-0.9$ for $0.15<M<0.6$ 
$M_{\sun}$, and $\alpha=-0.6$ for $0.02<M<0.15$ $M_{\sun}$.

We derived the expected $q$ distribution for our sample by 
assuming that every binary primary had a companion randomly 
drawn from the lower-mass regime of the CMF. Most 
implementations of this process use Monte Carlo simulations to 
draw a suitable population from the CMF, but our CMF is defined 
as a simple analytic function, so we chose to directly convert 
it into a $q$ distribution: 
$dN/dq=(dN/dM_{sec})(dM_{sec}/dq)$$\propto$$\Psi$$(q M_{prim})/M_{prim}$, 
where 
the full distribution $f(q)$ is the normalized sum of all 
functions $dN/dq$ as defined for each binary primary. In the 
three left-hand panels of Figure 6, we plot our IMF-based $q$ 
distribution with a red long-dashed line. Unlike the previous 
two distributions, our IMF-based distribution is fundamentally 
different for our entire sample and for each subsample since 
they represent different sets of primary masses.

The bottom-heavy nature of the IMF suggests that of all sources 
with masses $\la$1 $M_{\sun}$, approximately 1/4 should be 
substellar and many of the rest should fall at the very bottom 
of the stellar mass range. This distribution disagrees very 
significantly with our results, and all statistical tests 
conclusively rule out the possibility that the companions in 
our sample might have been randomly drawn from the IMF.

\subsubsection{A Parameterized Log-Normal Distribution}

As well as simply testing fixed distributions, we can use 
Bayesian analysis to draw conclusions about the most likely 
models from a class of distributions. We chose distributions 
that are log-normal in $q$ (base-10), with a mean at $q=1$. We 
chose this distribution because it is based on the following 
ad-hoc model: beginning with two equal-mass cores, we accrete 
matter stochastically onto the two cores such that the mean 
accretion rate onto each core is proportional to the core mass. 
Applying the central limit theorem to the logarithm of the core 
mass ratio, we arrive at a log-normal distribution in $q$. This 
distribution also has the important property that the 
functional form is the same in $1/q$ as in $q$, meaning that it 
does not matter whether the ``primary'' or ``secondary'' star is 
used as the reference for calculating $q$. Amongst 
differentiable $q$ distributions, only distributions that have 
an asymptotic power-law slope of -1 at $q=1$ can be written so 
that they have this property. This distribution has a 
corresponding probability density function:

\begin{equation}
 f(q) = \exp(-\log(q)^2/2 \sigma^2)/q.
\end{equation}

The likelihood function is then given by:

\begin{equation}
 L(\{q_i\} | \sigma) = \Pi_i \frac { 
\exp(-\log(q_i)^2/2\sigma^2 )/q_i }  
 {\int_{0.04}^{1.0} \exp(-\log(x)^2/2\sigma^2 )/x dx}.
\end{equation}

The normalization in this equation explicitly includes our 
lower limit for $q$. Using a uniform prior on $\sigma$, we find 
that the best fit value of $\sigma$ is 
$0.428^{+0.059}_{-0.049}$. This is our best fit distribution of 
all tested distributions, and predicts that only 1.2\% of all 
companions are brown dwarfs in our separation range (meaning $q 
< 0.08$ here). It reproduces the peak in the companion 
distribution at $q \sim 0.4$ similar to that seen by DM91, but 
without the lack of near equal-mass companions predicted by 
their preferred distribution. 

The low-mass subsample has a best fit value of $\sigma$ of 
$0.347^{+0.063}_{-0.049}$, and the high-mass subsample has a 
best fit $\sigma$ of $0.528^{+0.148}_{-0.092}$. These values of 
$\sigma$ are significantly different at the 96\% level. This 
demonstrates that the lower mass subsample prefers more 
equal-mass companions to the higher mass subsample, consistent 
with results for low-mass binaries in the field. We have plotted all 
three best-fitting log-normal distributions in the right-hand 
panels of Figure 6 (magenta dashed line).

\subsection{The Binary Separation Distribution}


Field surveys have also suggested that the separation 
distribution depends strongly on mass; the shape seems to be 
log-normal for a wide range of masses, but the mean and maximum 
separations decline with decreasing mass. DM91 found that the 
separation distribution for solar-mass stars has a mean 
separation of $\sim$30 AU and some binaries as wide as 10$^4$ 
AU. FM92 and RG97 found that early M binaries have a mean 
separation which is marginally consistent (4-30 AU), but few 
have separations $\ga$10$^3$ AU. Finally, recent surveys have 
shown that late M dwarfs and brown dwarfs have very small mean 
and maximum separations (4 and 20 AU, respectively; Close et 
al. 2003; Bouy et al. 2003; Burgasser et al. 2003). As we 
described above, many field stars formed in denser 
environments, so there has probably been some dynamical 
evolution that disrupted wide binaries. However, surveys of 
older clusters (e.g. Patience et al. 2002) suggest that the old 
binary population is only severely depleted by intra-cluster 
dynamical interactions at separations of $\ga$100-200 AU. This 
suggests that only the outer edge of our sample's separation 
distribution should differ significantly from the field.

Interpretation of the companion separation distribution is 
usually complicated by observational realities. The most 
meaningful quantity to consider is the distribution of 
semimajor axes, but the semimajor axis can only be determined 
as part of an orbital solution. Some authors convert the 
projected separation for each star into an estimated semimajor 
axis using a single corrective factor (typically $a=1.26r$), 
but this choice is only valid on a statistical level and 
carries implicit assumptions about the eccentricity 
distribution that are extrapolated from much shorter-period 
binaries. Therefore, we choose to report the observed projected 
separation distribution only. In the right side of Figure 6, we 
plot histograms of the separation distributions for our entire 
sample, only the higher-mass FGK stars, and only the lower-mass 
M dwarfs. This distribution spans separations of 6-435 AU, the 
range where our survey is sensitive to most brown dwarf companions.

We find that the separation distribution for our sample is 
consistent with a distribution constant with log($r$), with $r$ the 
apparent separation on the sky. A one-sided Kolmogorov-Smirnoff test 
over the separation range 6-435\,AU gives $D=0.13$, with $p=0.57$. 
In order to examine what our separation distribution is not 
consistent with, we have also attempted to fit log-normal 
distributions over the separation range 6-435 AU, where the 
likelihood of a particular value of the mean $\mu$ and standard 
deviation $\sigma$ is given by:

\begin{equation}
 L(\{r_i\} | \mu,\sigma) = \Pi_i \frac { \exp(-(\mu - \log(r_i))^2/2\sigma^2 ) }
 {\int_{\log (r_{\rm min})}^{\log(r_{\rm max})} \exp(-(\mu - x)^2/2\sigma^2) dx}.
\end{equation}

As in the previous subsection, the normalization on the denominator is an 
explicit integral rather than the standard normalization for a Gaussian 
because of our artificial truncation of the distribution at 6 and 435\,AU. 
We take the prior distribution of $\mu$ to be uniform between 0 and 3 
(i.e. median separations between 1 and 1000\,AU), and the prior 
distribution of $\sigma$ to be uniform between 0 and 2. The most likely 
values of $\mu$ and $\sigma$ are then 1.44 and 1.01 for the entire sample, 
1.08 and 0.79 for the high-mass sample, and 1.92 and 0.97 for the low-mass 
sample. However, integrating over all $\mu$, the most likely value of 
$\sigma$ is our upper limit of 2, demonstrating that the data is 
consistent with an approximately flat distribution. The most important 
point to come out of this analysis is that the 90\% confidence lower limit 
on $\sigma$ is 0.94, suggesting that we have detected at most two thirds of the 
companions in our sample, with the remaining companions being at smaller 
or greater separations.

The separation distributions for the high and low-mass samples follows the 
opposite trend to that suggested in the literature. Our low-mass sample 
has a median separation of 81\,AU, while our high-mass sample has a median 
separation of 21\,AU. This difference is not statistically significant 
since both distributions are consistent with a constant distribution, and 
a 2-sided Kolmogorov-Smirnoff test gives a difference statistic $D=0.30$, 
with $p=0.36$. It is interesting, however, that we do not see the trend 
toward smaller separations with lower masses as seen in field dwarfs (e.g. 
Allen (2007), who finds $\mu=0.86$ and $\sigma=0.28$ for ultracool field 
dwarfs). We hope to repeat this analysis with more conclusive results 
after we complete our survey sample.

\subsection{The Total Binary Fraction}

The total binary fraction, representing the integrated 
separation and mass ratio distributions, provides a useful 
comparison for different populations. It does not provide any 
additional information about the binary formation process that 
is not implicitly included in its component distributions, but 
it is very useful in other contexts like correcting the IMF for 
undetected multiplicity or relating the IMF to the prestellar 
core mass function. Previous surveys suggest that the binary 
fraction is close to unity for early-type stars, declining to 
$\sim$60\% for solar-mass stars, and $\sim$30\% for early M 
stars; in all cases, $\sim$40-50\% of binaries fall within the 
same separation range as our survey (6-435 AU).

We found binary fractions of 35$^{+5}$$_{-4}$\% for our entire 
sample, 33$^{+7}$$_{-5}$ for our high-mass (FGK) subsample, and 
38$^{+7}$$_{-6}$\% for our low-mass (early M) subsample. The first 
two results are roughly consistent with those observed in the field, 
but the second result is significantly higher than the value 
observed in the field. A survey of wide multiplicity has found that 
there are only four binaries with separations of 3-30\arcsec\, among 
our sample members (Kraus \& Hillenbrand 2007a; Kraus \& 
Hillenbrand, in preparation), but there are likely to be a 
significant number at smaller separations; we discovered some of 
these binaries inside the nominal completeness limit of our survey, 
and future RV surveys are likely to uncover many more. If the binary 
fraction at separations $\la$6 AU is as high in Upper Sco as in the 
field, then the binary fraction for early M dwarfs in Upper Sco 
could be as high as is observed for field F-G dwarfs ($\ga$60\%).

\section{The Farthest Shore?}

In the past 15 years, the search for extrasolar planets has become 
one of the major goals of the astronomical community. Radial 
velocity searches have discovered hundreds of planets and allowed us 
to probe the dynamics of planetary systems (e.g. Marcy et al. 2005), 
and more recently, transit searches have uncovered dozens of 
additional planets and allowed us to study their fundamental 
properties (masses and radii; O'Donovan et al. 2007; Torres 2007). 
However, the direct observation of extrasolar planetary systems has 
proven to be an elusive goal. Advances in high-resolution imaging 
(mostly aimed at speckle suppression) are allowing for increasingly 
strict upper limits on their existence, but no planetary companions 
at separations comparable to our own solar system have been directly 
imaged yet. An intriguing sample of candidate planetary-mass 
companions have been identified at much wider separations (e.g. 
Chauvin et al. 2004; Neuh\"auser et al. 2005), but their mass and 
formation mechanism are still uncertain.

The difficulty of directly detecting extrasolar planets with existing 
methods suggests that a change of strategy is in order. Previous surveys 
have typically used spectral or rotational differential imaging (Masciadri 
et al. 2005; Biller et al. 2007; Lafreniere et al. 2007) to cancel AO 
speckles, though some surveys have also used direct imaging (typically in 
the mid-infrared; Kasper et al. 2007) and simply accepted the inherent 
limits from speckle noise. All of these surveys produce their deepest 
limits at wide separations ($\ga$0.5\arcsec), so they can only probe the 
regime of likely planet formation (5-30 AU) for relatively nearby stars 
($d\la$30 pc); even for these stars, existing surveys can not probe deep 
enough to identify old ($\tau$$\ga$1 Gyr) planets, so they must study 
intermediate-age ($\tau$$\sim$10-200 Myr) members of nearby moving groups. 
By contrast, our survey achieves its deepest limits at much smaller 
angular separations, so we can probe deeper into the planetary separation 
regime of nearby moving group members (Ireland \& Kraus, in prep) and 
finally systematically survey the nearest very young associations like 
Upper Sco.

However, we must include a cautionary note: the fact that we found 
no high-confidence planetary detections could allow us to place 
upper limits on the existence of massive Jupiter analogues, but as 
we have previously described, it could also show that current models 
severely overestimate the luminosity of young planets. The 
core-accretion models which predicted this underluminosity have 
difficulty producing 10 $M_{Jup}$ planets, so it is possible that 
massive planets are formed via disk fragmentation (which may not 
suffer this underluminosity). However, all of our subsequent results 
should be taken with some skepticism. We list all of our detection 
limits in Table 6, so if the models are updated in the future, it 
should be trivial to re-analyze our results and produce new limits.

\subsection{Modeling the Population of Young Planets}

We expect that the planetary population over our range of interest 
will be described by three parameters: the total frequency $f$, a 
power-law mass distribution $dN/dM\propto$$M^{\alpha}$, and a 
power-law semimajor axis distribution $dN/da\propto$$a^{\beta}$. We 
can place constraints on these parameters by simulating a population 
of planetary systems for each set of parameters, then convolving 
this population with our detection limits to determine the level of 
consistency with our nondetection. Our survey's detection limits 
can't be directly translated into limits on the planetary population 
since planets could be obscured by projection effects, so for each 
simulated planet, we also invoke a random inclination angle, a 
random true anomaly, and an eccentricity drawn from the 
approximately Gaussian distribution observed for radial velocity 
planets (Juric \& Tremaine 2007). We note that Juric \& Tremaine 
chose to fit their eccentricity distribution with a Schwartzschild 
function, but given the uncertainties in the observational 
statistics, it is not possible to determine whether a Schwartzschild 
or Gaussian function is more appropriate. We have adopted the more 
computationally convenient form. 

Our specific implementation uses a mass drawn from between 1 and 30 
$M_{Jup}$, a semimajor axis drawn from between 3 and 36 AU, and an 
eccentricity drawn from a Gaussian distribution between 0.0 and 0.8 
with mean $\mu$$_e=0.3$ and standard deviation $\sigma$$_e=0.3$. We 
do not directly model the planetary frequency $f$ in our Monte Carlo 
routine because it can be added analytically. We adopted the upper 
mass limit (30 $M_{Jup}$) to match the most massive T Tauri disks at 
ages of 1-2 Myr (only $\sim$1\% of which significantly exceed this 
mass; Andrews \& Williams 2005). After conducting our simulations 
for a range of values of $f$, $\alpha$, and $\beta$, we compiled a 
three-dimensional probability density function $P(f,\alpha,\beta)$ 
which corresponds to the probability that we would have detected a 
planet, then extracted three-dimensional confidence surfaces which 
correspond to the 50\%, 90\%, 95\%, and 99\% probabilities that our 
observations actually would have found no planets.

\begin{figure}
\plotone{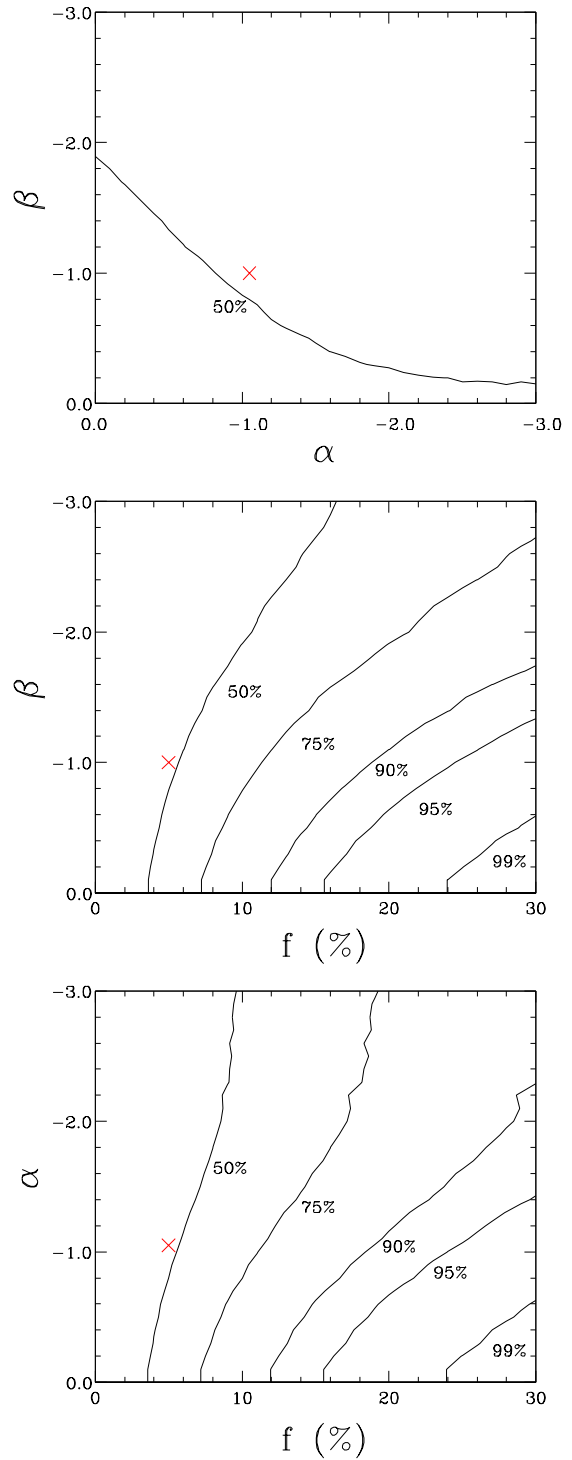}
\caption{Our survey's joint limits on the total giant planet frequency 
$f$, the mass function power law $\alpha$, and the semi-major axis 
distribution power law $\beta$, assuming we fix each parameter at the 
canonical value suggested by RV surveys (e.g. Marcy et al. 2005): $f=5\%$ 
(top), $\alpha=-1.05$ (middle), and $\beta=-1.0$ (bottom). In each case, 
we also denote the confidence level corresponding to all three canonical 
values with red crosses.}
\end{figure}

\subsection{Limits on the Population of Young Planets}

It is difficult to present a set of three-dimensional confidence surfaces 
in a two-dimensional medium, so we have chosen to present a selection of 
two-dimensional slices where we fix one parameter to its current 
best-estimated value. The statistics of radial velocity surveys have 
finally become significant enough to suggest possible values of our 
distribution parameters, so we have adopted these canonical values 
($f=5\%$, $\alpha=-1.05$, $\beta=-1.0$; Marcy et al. 2005) to produce our 
three confidence plots. The canonical distribution values are derived from 
radial velocity surveys; they have found the power-law exponents $\alpha$ 
and $\beta$ for their sample of (short-period) planets, and they 
extrapolate that $\sim$5\% of their sample members have long-term linear 
RV trends suggestive of massive long-period planets. The power law 
exponents may not be valid since many gas giants at small separations are 
thought to have migrated there, but these values represent the best 
constraint available.

In the three panels of Figure 7, we present the joint confidence intervals 
for each pair of values if we fix the third value to the canonical 
estimate. These results suggest that the canonical planetary distribution 
can only be ruled out at the $\sim$50\% level. This is not a statistically 
significant level, but it is much better than any previous imaging survey 
could have achieved. We also find that a much higher planetary frequency 
is significantly ruled out for most values of $\alpha$ and $\beta$; the 
only values which are consistent require either the mass function or the 
separation distribution to be very steep, placing most planets in a regime 
that our survey can not search. Otherwise, we can not rule out significant 
regions of parameter space. In particular, if the canonical planetary 
frequency ($f=$5\%) is accurate, then we can not place any constraints 
beyond the 70\% level on values of $\alpha$ or $\beta$. This is a 
straightforward result of our sample size; with 60 targets, a frequency of 
5\% suggests that only 3 wide planets exist in our sample. Only 
unrealistically top-heavy mass or separation distributions would place a 
significant number of planets in our survey's detection limits.

Finally, we can determine a direct constraint on the total frequency of 
wide high-mass planets by adopting the canonical values for both $\alpha$ 
and $\beta$, reducing the confidence surface to a confidence interval. If 
we assume that $\alpha$$=-1.05$ and $\beta$$=-1.0$, then there is a 90\% 
probability that $f<19\%$ and a 95\% probability that $f<24\%$. We can not 
place similar limits on $\alpha$ and $\beta$ because these confidence 
limits correspond to a total of 2.3 and 3.0 expected detections, 
respectively; if the total planetary frequency is only $f=5\%$, then only 
extremely top-heavy power laws would allow for that many expected 
detections among our 60 targets.

\subsection{An Ocean in the Distance?}

Our survey did not identify any faint companions at a confidence level of 
$\ga$99.5\%, but it did identify four faint candidate companions at 
confidence levels of 97.5\% to 99.5\%. Based on our total sample size (60 
targets), the expected number of spurious detections with a confidence 
level $>$97.5\% is only $\sim$1.5; according to Poisson statistics, the 
probability of identifying 4 or more of these spurious detections is only 
7\%, so this seems to represent a marginally significant excess.  We will 
try to obtain followup observations for each marginal detection in the 
upcoming observing season; given their faintness, any genuine companion 
in this group could represent the first directly-imaged massive 
Jupiter analogue.

\section{How Arid is the Brown Dwarf Desert?}

Many multiplicity surveys suggest that the binary companion mass 
function declines as it enters the brown dwarf mass range, and 
all results from RV surveys suggest the same for the planetary 
mass function. In light of these results, it is not surprising 
that brown dwarf companions are uncommon. The interesting 
question is whether they are more uncommon than predicted by 
the tails of both mass functions; if so, then this deficit 
genuinely represents a brown dwarf ``desert''.

Our results suggest that the stellar mass ratio distribution is 
constant or at least not biased heavily toward low-mass 
companions. Given our observed total binary fraction 
(35$^{+5}_{-4}$\%), a constant mass ratio distribution predicts 
that $\sim$3.5\% of all stars should have a substellar or 
nearly-substellar companion with $q\le0.1$ at separations of 
6-435 AU. We have found two such companions 
(1.8$^{+2.3}_{-0.6}$\%), a result which is entirely consistent 
with that claim. However, both companions fall at the upper end 
of this range ($q=0.10$ and $q=0.09$), and only one is possibly 
substellar ($M_{sec}\sim$0.07 $M_{\sun}$); given the 
uncertainties inherent to our estimates of stellar properties, 
it is not inconceivable that both companions could fall in the 
range $q>0.1$. This would be consistent with estimates for wide 
companions to much higher-mass Upper Sco members; Kouwenhoven 
et al. (2007) found that only 0.5$\pm$0.5\% of the B and A 
stars in Upper Sco have substellar companions with separations 
of 130-520 AU.

Our estimate of the contribution from planetary formation 
processes is much more uncertain. If the planetary distribution 
is truly defined by the canonical values given in the 
literature ($f=5\%$, $\alpha$$=-1.05$, and $\beta$$=-1.0$) and 
the models describing luminosities of young planets are 
correct, then our survey would have had a 50\% chance of 
detecting one ``planetary'' companion of any mass $<$30 
$M_{Jup}$. This probability would have been higher if the 
planetary mass function extends beyond 30 $M_{Jup}$ with no 
cutoff, but even a cutoff at 100 $M_{Jup}$ would imply 
that our null detection is significant at only $\sim$75\%. As a 
result, we can not state with any confidence that the canonical 
values are incorrect or that there is any sort of high-mass 
cutoff in the planetary mass function.

\section{Summary}

We present the results of a survey for stellar and substellar companions 
to 82 young stars in the nearby OB association Upper Scorpius. This survey 
used nonredundant aperture-mask interferometry to achieve typical contrast 
limits of $\Delta$$K\sim$5-6 at the diffraction limit, revealing 12 new 
companions that lay below the detection limits of traditional 
high-resolution imaging; we also summarize a complementary snapshot 
imaging survey that discovered 7 directly resolved companions. The overall 
frequency of binary companions ($\sim$33$^{+5}_{-4}$\% at separations of 
6-435 AU, including companions reported in the literature) appears to be 
similar to field stars of similar mass, but the companion mass function 
appears to be more biased toward equal-mass companions than the equivalent 
mass function in the field. This result could indicate an environmental or 
dynamical effect, but our number statistics are not yet sufficient to 
place strong constraints on its nature.

Our survey limits encompass the entire brown dwarf mass range and we 
detected two companions with $q\le$0.1, a number which is consistent 
with a flat mass ratio distribution. However, both of these 
companions have mass ratios near 0.1 and only one has a mass which 
might fall below the substellar boundary, so we hesitate to rule out 
the existence of any deficit that might denote a brown dwarf 
``desert''. Our survey's deep detection limits also extend into the 
top of the planetary mass function; we have not identified any 
planetary companions at high confidence ($\ga$99.5\%), but we have 
identified four candidate companions at lower confidence 
($\ga$97.5\%) that merit additional followup to confirm or disprove 
their existence. The lack of planets within the brown dwarf mass 
range also is not a significant proof of the existence of a desert.

Finally, we note that our survey results are extremely encouraging with 
respect to the potential for future discoveries. We are currently 
extending our survey efforts to the Taurus-Auriga star forming region and 
to several nearby moving groups, and this expansion of our sample should 
make any conclusions much more robust. Our ability to precisely measure 
astrometry for close ($\sim$2-3 AU) binary systems could also allow us to 
measure dynamical masses for many young stars on a timescale of $\la$5 yr. 
Finally, achieving similar detection limits for planetary-mass companions 
in Taurus-Auriga and the nearby moving groups will significantly enhance 
our limits on the properties of young planets; a similar null detection 
for our full sample would significantly rule out the canonical values for 
the planetary distribution function, confirming either that these values 
are wrong or that evolutionary models significantly overestimate the 
luminosity (and detectability) of young planets.

\section{Appendix A: The Detection Limits of Nonredundant Aperture Mask Observations}

For each set of $n$ frames (called a {\em run}), with $n \ge 8$, we
calculate the mean closure phase vector $\bar{\phi}$ and an
estimate for the covariance matrix of 
closure-phase:

\begin{equation}
 \hat{C_r} = \frac{ \Sigma_i (\phi_i - \bar{\phi})^t (\phi_i - \bar{\phi}) }
 {n-1}.
\end{equation}

Here $\phi_i$ is the closure-phase vector calculated for a single frame 
and $^t$ represents a transpose. The variance in closure-phase calculated 
by this technique (the diagonal of $\hat{C_r}$) will be called 
$\hat{\sigma^2_r}$.

Studentizing all statistics was seen as an excessively difficult task,
given the high number of dimensions in our data set, the strong
correlations between measured parameters and the need to have fast,
automatic fitting routines. In order to limit susceptibility to the
lack of a tail in the Gaussian approximation for uncertainties that
follow a student's $t$ distribution, we
artificially increased the errors on closure phases with the smallest
errors. We did this by applying an error cutoff at 2/3 of the median
closure-phase error. In the case of closure-phases with equal true
errors, this means that we would have artificially increased the
uncertainties on 12\% of the closure triangles, changing the expected
value of reduced $\chi^2$ from 1.4 to 1.24 for $n=8$, making a smaller
difference for large $n$. 

In addition to the error calculated from a single run, we calculated
closure-phase uncertainties from the dispersion amongst calibrator
observations. We will denote these variances $\hat{\sigma^2_c}$. Where
the error calculated from dispersion amongst the calibrators was
greater than that given by the standard error of the mean for a single
suns, we weighted the error estimates by: 

  \begin{equation}
    \hat{\sigma^2} = \frac{2 \hat{\sigma_r^2} + (n_c - 1) \hat{\sigma_c^2}}
      {n_c + 1}
  \end{equation}

After errors were increased by either the scatter amongst calibrators
or the closure-phase uncertainty histogram cutoff, the covariance
matrix was modified in such a way that the correlation matrix
remained unchanged. 

After finding the covariance matrix of calibrated closure-phase, we
found that in general the errors in its calculation caused excessive
noise in calculation of the covariance matrix inverse. For this
reason, we first filtered the covariance matrix by fitting a model of the
form:

\begin{equation}
 {\rm Var}(\theta_{ijk}) = \frac{\sigma^2_f}{\langle |V_i V_j
 V_k|\rangle} + {\rm Var}(\theta_i) + {\rm Var}(\theta_j) + {\rm Var}(\theta_k),
\end{equation}

and

\begin{equation}
  {\rm Cov}(\theta_{ijk},\theta_{jlm}) = \pm Var(\theta_i).
\label{eqnCovCP}
\end{equation}

Var$(\theta_i)$ is a model variance of phase for baseline $i$, which
in turn has the form:

\begin{equation}
 {\rm Var}(\theta_i) = \sigma^2_i + \alpha(m_{tx} m_{tx}^t + m_{ty} m_{ty}^t).
\end{equation}

Here $\sigma_i$ is the intrinsic phase variance of baseline $i$,
$\alpha$ is a free parameter and $m_{tx}$ and $m_{ty}$ are
closure-phase modes caused by skewness of the image in $x$ and $y$
directions. This skewness is caused by temporal effects, where in a
single exposure, tip/tilt errors can be asymmetrical, with e.g. a
single ``glitch'' where for 10\% of the exposure the image is offset
by 20\,mas. This can be a dominant error term at Palomar, where the
tip/tilt mirror is too large to have an adequate correction bandwidth
in poor seeing. The $\pm$ in Equation~\ref{eqnCovCP} is a $+$ if baseline $i$ is
counted in the same direction for both closure-phases, and it is a $-$
if baseline $i$ is counted in opposite directions for both
closure-phases. 

Finally, in the fitting process, reduced $\chi^2$ was often greater
than 1. Although by chance, this should have occured 50\,\% of the time if
uncertainties were correctly estimated, in practice it occured $\sim$90\,\%
of the time. A possible reason for this could be residual systematics
differences in sky position or color, despite the care taken to
minimise these differences in a single observing block. When this
occured, additional systematic closure-phase uncertainties were added so
that reduced $\chi^2$ was 1. For determining the confidence level of a
null detection, this reduced $\chi^2$ corresponds to the reduced
$\chi^2$ for a single star fit, and for determining errors on a
non-null detection, this reduced $\chi^2$ corresponds to that for the
best binary fit.

Due to the linear dependence of model closure-phases, we calculate
$\chi^2$ on a closure-phase vector space with dimensionality equal to
the number of independent closure phases, $N_{\rm ind}$. This vector
space $V_{\rm ind}$ is formed by projection via a matrix $T_p$,
defined so that the covariance matrix on $V_{\rm ind}$ is a diagonal
matrix $D$:

\begin{equation}
 D = T_p C^{-1} T_p^t.
\end{equation}

This means that, given a closure phase vector ${\bf \phi_d}$, the
covariance matrix of the linear combination $T_p {\bf \phi_d}$ is
given by the diagonal matrix D. Given model closure phases 
${\bf \phi_m}$ and data ${\bf \phi_d}$, the value of $\chi^2$ is then
given by: 

\begin{equation}
 \chi^2 = ({\bf \phi_m-\phi_d})^t T_p^t D T_p ({\bf \phi_m-\phi_d})
\end{equation}

It is this $\chi^2$ that was minimised when fitting binary functions
to closure-phase. The Monte-Carlo procedure was simplified
computationally by limiting 
the fitting prodecure to the high-contrast regime, where closure-phase
is a linear function of companion brightness. In this regime, model
closure-phases ${\bf \phi_m}$ were found for each separation and position
angle for a fixed contrast ratio $R_m$, with the fitted contrast
ratio given by:

\begin{equation}
 R = R_m \frac{     {\bf \sigma Z^t}   D T_p {\bf \phi_m} }
              { {\bf \phi_m^t} T_p^t D T_p {\bf \phi_m} } 
\end{equation}

The matrix ${\bf \sigma Z}$ is a standard normal
vector of length $N_{\rm ind}$, multiplied by the standard errors as
calculated in the vector space $V_{\rm ind}$. This equation is
relatively simple to derive my minimizing $\chi^2$ where the model closure
phase at contrast $R$ is $(R/R_m) {\bf \phi_m}$.

\acknowledgements

The authors thank Lynne Hillenbrand for helpful feedback on the 
manuscript and Brian Cameron for sharing his NIRC2 astrometric 
calibration results prior to publication. We also thank the referee, 
Ralph Neuh\"auser, for providing a helpful critique. We recognize 
and acknowledge the very significant cultural role and reverence 
that the summit of Mauna Kea has always had within the indigenous 
Hawaiian community. We are most fortunate to have the opportunity to 
conduct observations from this mountain.

M.I. would like to acknowledge Michelson Fellowship support from the 
Michelson Science Center and the NASA Navigator Program. This work 
is also partially supported by the National Science Foundation under 
Grant Numbers 0506588 and 0705085. This work makes use of data 
products from 2MASS, which is a joint project of the University of 
Massachusetts and IPAC/Caltech, funded by NASA and the NSF. Our 
research has also made use of the USNOFS Image and Catalogue Archive 
operated by the United States Naval Observatory, Flagstaff Station 
(http://www.nofs.navy.mil/data/fchpix/).


\begin{thebibliography}{}
\bibitem[Ahmic et al.(2007)]{ahm07} Ahmic, M., Jayawardhana, R., 
Brandeker, A., Scholz, A., van Kerkwijk, M., Delgado-Donate, E., \& 
Froebrich, D. 2007, \apj, 671, 2074
\bibitem[Andrews \& Williams(2005)]{aw05} Andrews, S. \& Williams, J. 
2005, \apj, 631, 1134
\bibitem[Ardila et al.(2000)]{ard00} Ardila, D. et al. 2000, \aj, 120,
  479
\bibitem[Allen (2007)]{allen07} Allen, P. 2007, \apj, in press, 
(arXiv:0707.2064)
\bibitem[Baraffe et al.(1998)]{bar98} Baraffe, I., Chabrier, G., Allard, 
F., \& Hauschildt, P. 1998, \aap, 337, 403
\bibitem[Bate \& Bonnell(2005)]{bb05} Bate, M. \& Bonnell, I. 2005, 
\mnras, 356, 1201
\bibitem[Bessell \& Brett(1988)]{bb88} Bessell, M. \& Brett, J. 1988, 
\pasp, 100, 1134
\bibitem[Biller et al.(2007)]{bil07} Biller, B. et al. 2007, \apjs, in 
press
\bibitem[Bouy et al.(2003)]{bouy03} Bouy, H. et al. 2003, \aj, 126, 1526
\bibitem[Bouy et al.(2006)]{bouy06} Bouy, H. et al. 2006, \aap, 451, 177
\bibitem[Brandner et al.(1996)]{b96} Brandner, W. et al. 1996, \aap, 307, 
121
\bibitem[Burgasser et al.(2003)]{burg03} Burgasser, A. et al. 2003, \apj, 
125, 850
\bibitem[Carpenter(2001)]{c01} Carpenter, J. 2001, \aj, 121, 2851
\bibitem[Carson et al.(2005)]{car05} Carson, J. et al. 2005, AJ, 130, 1212
\bibitem[Chabrier et al.(2000)]{cha00} Chabrier, G. et al. 2000, ApJ, 542, 464
\bibitem[Chauvin et al.(2004)]{ch04} Chauvin, G., Lagrange, A., Dumas, 
C., Zuckerman, B., Mouillet, D., Song, I., Beuzit, J., \& Lowrance, P. 
2004, \aap, 425, 29
\bibitem[Chiu et~al.(2006)]{Chiu06}
{Chiu}, K., {Fan}, X., {Leggett}, S.~K., {Golimowski}, D.~A., {Zheng}, W.,
  {Geballe}, T.~R., {Schneider}, D.~P., \& {Brinkmann}, J. 2006, \aj,
  131, 2722
\bibitem[Close et al.(2003)]{c03} Close, L. et al. 2003, \apj, 587, 407
\bibitem[de Zeeuw et al.(1999)]{dZ99} de Zeeuw, P., Hoogerwerf, R., de 
Bruijne, J., Brown, A., \& Blaauw, A. 1999, \aj, 117, 354
\bibitem[Delgado-Donate et al.(2003)]{dd03} Delgado-Donate, E. et al. 
2003, \mnras, 347, 759
\bibitem[Duquennoy \& Mayor(1991)]{dm91} Duquennoy, A. \& Mayor, M. 
1991, \aap, 248, 485
\bibitem[Fischer \& Marcy(1992)]{fm92} Fischer, D. \& Marcy, G. 1993, 
\apj, 396, 178
\bibitem[Gizis et al.(2001)]{giz01} Gizis, J., Kirkpatrick, J.D., 
Burgasser, A., Reid, I.N., Monet, D., Liebert, J., \& Wilson, J. 
2001, \apj, 551, L163
\bibitem[Goodwin et al.(2007)]{good07} Goodwin, S. \& Kroupa, P. 2007, in 
Protostars and Planets V, ed. B. Reipurth, D. Jewitt, \& K. Keil (Tucson: 
Univ. Arizona Press), 133
\bibitem[Grether \& Lineweaver(2006)]{gl06} Grether, D. \& Lineweaver, C. 2006, ApJ, 640, 1051
\bibitem[Hillenbrand \& White(2004)]{hw04} Hillenbrand, L. \& White, R. 
2004, \apj, 604, 741
\bibitem[Houk \& Smith-Moore(1988)]{hsm88} Houk, N. \& Smith-Moore, M. 
1988, Michigan Spectral Survey (Ann Arbor: Dept. Astron. UNiv. Mich.),4 
\bibitem[Hubber \& Whitworth(2005)]{hw05} Hubber, D. \& Whitworth, A. 
2005, \aap, 437, 113
\bibitem[Johnson et al.(2006)]{john06} Johnson, J. et al. 2006, \apj, 
647, 600
\bibitem[Juric \& Tremaine(2007)]{jt07} Juric, M. \& Tremaine, S. 2007, 
submitted to ApJ (astro-ph/0703160)
\bibitem[Kim et al.(2005)]{kim05} Kim, S., Figer, D., Lee, M., \& Oh, S. 
2005, \pasp, 117, 445
\bibitem[Kirkpatrick et~al.(2000)]{Kirkpatrick00}
{Kirkpatrick}, J.~D., {Reid}, I.~N., {Liebert}, J., {Gizis}, J.~E.,
  {Burgasser}, A.~J., {Monet}, D.~G., {Dahn}, C.~C., {Nelson}, B., \&
  {Williams}, R.~J. 2000, \aj, 120, 447
\bibitem[K\"ohler et al.(2000)]{koh00} K\"ohler, R. et al. 2000, 
\aap, 356, 541
\bibitem[Kouwenhoven et al.(2007)]{kou07} Kouwenhoven, M.B.N., 
Brown, A.G.A., \& Kaper, L. 2007, \aap, 464, 581
\bibitem[Kraus et al.(2006)]{kraus06} Kraus, A., White, R., \& 
Hillenbrand, L. 2006, \apj, 649, 306
\bibitem[Kraus \& Hillenbrand(2007a)]{kraus07} Kraus, A. \& Hillenbrand, 
L. 2007, \apj, 662, 413
\bibitem[Kraus \& Hillenbrand(2007b)]{kraus07b} Kraus, A. \& Hillenbrand, 
L. 2007, \apj, 664, 1167
\bibitem[Kroupa(1995)]{kr95} Kroupa, P. 1995, \mnras, 277, 1491
\bibitem[Kroupa \& Bouvier(2003)]{kb03} Kroupa, P. \& Bouvier, J. 2003, 
\mnras, 346, 369
\bibitem[Kroupa et al.(2003)]{krou03} Kroupa, P. et al. 2003, \mnras, 346, 
354
\bibitem[Kunkel(1999)]{kun99} Kunkel, M. 1999, PhDT
\bibitem[Ireland et al.(2008)]{ire08} Ireland, M., Kraus, A., 
Martinache, F., Lloyd, J.P., \& Tuthill, P.G. 2008, ApJ, in press 
(arXiv:0801.1525)
\bibitem[Lafreniere et al.(2007)]{laf07} Lafreniere, D. et al. 2007, 
\apj, in press
\bibitem[Leggett et al.(1998)]{leg98} Leggett, S., Allard, F., \& 
Hauschildt, P. 1998, \apj, 509, 836
\bibitem[Lissauer \& Stevenson(2007)]{lis07} Lissauer, J. \& Stevenson, 
D. 2007, in Protostars and Planets V, ed. B. Reipurth, D. JEwitt, \& K. 
Keil (Tucson: Univ. Arizona Press), 591
\bibitem[Lloyd et al.(2006)]{ll06} Lloyd, J., Martinache, F., Ireland, 
M., Monnier, J., Pravdo, S., Shaklan, S., \& Tuthill, P. 2006, ApJL, 650, 
131
\bibitem[{{Lohmann} {et~al.}(1983){Lohmann}, {Weigelt}, \&
  {Wirnitzer}}]{Lohmann83}
{Lohmann}, A.~W., {Weigelt}, G., \& {Wirnitzer}, B. 1983, \ao, 22, 4028
\bibitem[Luhman et al.(2003)]{luh03} Luhman, K. et al. 2003, \apj, 593, 
1093
\bibitem[Luhman(2004)]{luh04} Luhman, K. 2004, \apj, 602, 816
\bibitem[Luhman(2006)]{luh06} Luhman, K. 2006, \apj, 645, 676
\bibitem[McCarthy \& Zuckerman(2004)]{McCarthy04}
{McCarthy}, C., \& {Zuckerman}, B. 2004, \aj, 127, 2871
\bibitem[Marcy \& Butler(2000)]{mar00} Marcy, G. \& Butler, R. 2000, 
\pasp, 112, 137
\bibitem[Marcy et al.(2005)]{mar05} Marcy, G., Butler, R., Fischer, D., 
Vogt, S., Wright, J., Tinney, C., \& Jones, H. 2005, Prog. Theor. Phys. 
Suppl., 158, 24
\bibitem[Marley et al.(2007)]{mar07} Marley, M., Fortney, J., Hubickyj, 
O., Bodenheimer, P., \& Lissauer, J. 2007, \apj, 655, 541
\bibitem[Martin et al.(2004)]{mar04} Martin, E. et al. 2004, \aj, 127, 449
\bibitem[Martinache et al.(2007)]{mart07} Martinache, F., Lloyd, J., 
Ireland, M., Yamada, R., \& Tuthill, P. 2007, \apj, 661, 496
\bibitem[Masciadri et al.(20075)]{mas07} Masciadri, E., Mundt, R., 
Henning, Th., Alvarez, C., \&  Barrado y Navascues, D. 2005, \apj, 625, 
1004
\bibitem[Metchev(2005)]{met05} Metchev, S. 2005, PhD thesis(http://etd.caltech.edu/etd/available/etd-08262005-160055/)
\bibitem[Monet et al.(2003)]{monet03} Monet, D. et al. 2003, \aj, 125, 984
\bibitem[Naef et al.(2007)]{naef07} Naef, D. et al. 2007, \aap, 470, 721
\bibitem[Nakajima et al.(1989)]{nak89} Nakajima, T. et al. 1989, AJ, 97, 
1510
\bibitem[Neuh\"auser et al.(2003)]{neu03} Neuh\"auser, R., Guenther, 
E., Alves, J., Hu\'elamo, N., Ott, Th., \& Eckart, A. 2003, AN, 324, 
535
\bibitem[Neuh\"auser \& Guenther(2004)]{neu04} Neuh\"auser, R. \& 
Guenther, E. 2004, \aap, 420, 647
\bibitem[Neuh\"auser et al.(2005)]{neu05} Neuh\"auser, R., Guenther, 
E., Wuchterl, G., Mugrauer, M., Bedalov, A., \& Hauschildt, P. 2005, 
\aap, 435, 13
\bibitem[O'Donovan et al.(2007)]{od07} O'Donovan, F. et al. 2007, ApJ, 
663, 37
\bibitem[Patience et al.(2002)]{pat02} Patience, J. et al. 2002, \aj, 123, 
1570
\bibitem[Pravdo et al.(2006)]{pra06} Pravdo, S. et al. 2006, ApJ, 649, 389
\bibitem[Preibisch et al.(1998)]{pre98} Preibisch, T. et al. 1998, \aap, 
333, 619
\bibitem[Preibisch et al.(2001)]{pre01} Preibisch, T., Guenther, E., \& 
Zinnecker, H. 2001, \aj, 121, 1040
\bibitem[Preibisch et al.(2002)]{pre02} Preibisch, T., Brown, A., 
Bridges, T., Guenther, E., \& Zinnecker, H. 2002, \aj, 124, 404
\bibitem[Reid \& Gizis(1997)]{rg97} Reid, I. \& Gizis, J. 1997, \aj, 113, 2246
\bibitem[Reid et al.(2001)]{r01} Reid, I., Gizis, J., Kirkpatrick, J., \& 
Koerner, D. 2001, \aj, 121, 489
\bibitem[Schmidt-Kaler(1982)]{sk82} Schmidt-Kaler, Th., "Physical 
Parameters of the Stars", Landolt-Bornstein Numerical Data and 
Functional Relationships in Science and Technology, New Series, 
Group VI, Volume 2b, Springer-Verlag, Berlin, 1982
\bibitem[Scholz et al.(2006)]{sch06} Scholz, A., Jayawardhana, R., \& 
Wood, K. 2006, \apj, 645, 1498
\bibitem[Skrutskie et al.(2006)]{skr06} Skrutskie, M. et al. 2006, \aj, 
131, 1163
\bibitem[Slesnick et al.(2006a)]{sles06a} Slesnick, C., Carpenter, J., \& 
Hillenbrand, L.2006a, \aj, 131, 3016
\bibitem[Slesnick et al.(2006b)]{sles06b} Slesnick, C., Carpenter, J., 
Hillenbrand, L., \& Mamajek, E. 2006b, \aj, 132, 2665
\bibitem[Slesnick(2007)]{sles07} Slesnick, C. PhD thesis
\bibitem[Stephenson(1986)]{steph86} Stephenson, C.B. 1987, \apj, 300, 779
\bibitem[Sterzik \& Durisen(1998)]{sd98} Sterzik, M. \& Durisen, R. 1998, 
\aap, 339, 95
\bibitem[Stetson(1987)]{stet87} Stetson, P. 1987, \pasp, 99, 191
\bibitem[Torres(2007)]{tor07} Torres, G. 2007, ApJL, in press
\bibitem[Tuthill et al.(2000)]{tut00}Tuthill, P. et al. 2000, PASP, 112, 555
\bibitem[Walter et al.(1994)]{wal94} Walter, F. et al. 1994, \aj, 107, 692
\bibitem[Weinberg et al.(1987)]{wein87} Weinberg, M., Shapiro, S., \& 
Wasserman, I. 1987, \apj, 312, 367
\bibitem[White \& Basri(2003)]{wb03} White, R. \& Basri, G. 2003, \apj, 
582, 1109
\bibitem[Whitworth \& Stamatellos(2006)]{ws06} Whitworth, A. \& 
Stamatellos, D. 2006, \aap, 458, 817
\end{thebibliography}
\end{document}